\documentclass{article}

\usepackage{microtype}
\usepackage{graphicx}

\usepackage{booktabs} % for professional tables
\usepackage{placeins}
\usepackage{hyperref}

\usepackage[preprint]{icml2026}

\usepackage{amsmath}
\usepackage{amssymb}
\usepackage{mathtools}
\usepackage{amsthm}

\usepackage{listings}
\usepackage{xcolor}

\usepackage{tikz}
\usepackage{pifont}  % for \ding symbols (checkmark, cross)

\lstdefinelanguage{Prompt}{
  morekeywords={Given, Identify, Generate, For, The},
  sensitive=false
}

\lstdefinelanguage{diff}{
    basicstyle=\ttfamily\small,          % Font style
    morecomment=[f][\color{red!60!black}]-,   % - red
    morecomment=[f][\color{green!40!black}]+, % + green
    morecomment=[f][\color{blue}]{@@},        % @@ blue
    morecomment=[f][\color{gray}]{diff},      % diff grey
    morecomment=[f][\color{gray}]{index},
    keepspaces=true,
    showstringspaces=false,
    frame=single,                   % Adds a simple rectangular frame
    rulecolor=\color{black!30},     % Frame border color (Light Gray)
    framesep=4pt,                   % Padding between code and frame
    framerule=0.5pt,                % Thickness of the border line
    breaklines=true,                % Enable automatic line wrapping
    breakatwhitespace=false,        % Allow wrapping at any character
    postbreak=\mbox{\textcolor{gray}{$\hookrightarrow$}\space}, % Visual cue for wrapped lines
    aboveskip=0.8em,                % Space above the block
    belowskip=0.8em                 % Space below the block
}

\lstdefinestyle{promptstyle}{
  language=Prompt,
  basicstyle=\ttfamily\small,
  keywordstyle=\bfseries,
  commentstyle=\itshape\color{gray},
  frame=single,
  rulecolor=\color{black!40},
  numbers=none,
  breaklines=true,
  breakatwhitespace=true,
  showstringspaces=false,
  columns=fullflexible,
  keepspaces=true,
  xleftmargin=1em,
  xrightmargin=1em,
  aboveskip=0.6em,
  belowskip=0.6em
}

\usepackage[capitalize,noabbrev]{cleveref}

\theoremstyle{plain}

\theoremstyle{definition}

\theoremstyle{remark}

\usepackage{multirow}
\icmltitlerunning{Adversarial Benchmark Strengthening}

\begin{document}

\twocolumn[
\icmltitle{SWE-ABS: Adversarial Benchmark Strengthening Exposes Inflated Success Rates on Test-based Benchmark}

% \begin{icmlauthorlist}
% \icmlauthor{Your Name}{yyy}
% \end{icmlauthorlist}

% \icmlaffiliation{yyy}{Your Affiliation}

% \icmlcorrespondingauthor{Your Name}{your.email@example.com}

\icmlsetsymbol{equal_first}{$*$}
\icmlsetsymbol{corr}{$\dag$}

\begin{icmlauthorlist}
\icmlauthor{Boxi Yu}{equal_first,yyy}
\icmlauthor{Yang Cao}{equal_first,per}
\icmlauthor{Yuzhong Zhang}{cuhksz}
\icmlauthor{Liting Lin}{yyy}
\icmlauthor{Junjielong Xu}{cuhksz}
\icmlauthor{Zhiqing Zhong}{cuhksz}
\icmlauthor{Qinghua Xu}{yyy}
\icmlauthor{Guancheng Wang}{yyy}
\icmlauthor{Jialun Cao}{hkust}
\icmlauthor{Shing-Chi Cheung}{hkust}
\icmlauthor{Pinjia He}{corr,cuhksz}
\icmlauthor{Lionel Briand}{yyy,ottawa}
\end{icmlauthorlist}

\icmlaffiliation{yyy}{Lero, the Research Ireland Centre for Software, University of Limerick}
\icmlaffiliation{per}{Independent Researcher}
\icmlaffiliation{cuhksz}{The Chinese University of Hong Kong, Shenzhen }
\icmlaffiliation{hkust}{Hong Kong University of Science and Technology}
\icmlaffiliation{ottawa}{University of Ottawa}

\icmlcorrespondingauthor{Pinjia He}{hepinjia@cuhk.edu.cn}

\icmlkeywords{Benchmark Evaluation, Adversarial Testing, Test Suite Quality, Program Repair, LLM Evaluation}

\vskip 0.3in
]

% \printAffiliationsAndNotice{} % empty if not using equal contribution
\printAffiliationsAndNotice{ {$*$} Equal contribution. {$\dag$} Corresponding author.}
% -----------------------------
% Main content via input files
% -----------------------------
% use the new abstract since it requires 4 to 6 sentences
\begin{abstract}

% base version-1
% The top-performing system on the SWE-Bench Verified leaderboard achieves a 78.80\% success rate, suggesting that the benchmark is approaching saturation.
% However, our re-evaluation of patches produced by the top-30 leaderboard agents reveals that 19.78\% of cases labeled as ``solved'' are semantically incorrect, passing only because limitations in the test suites fail to expose their errors.
% To address this, we present SWE-ABS, an adversarial framework that strengthens test suites through a two-stage pipeline: coverage-driven augmentation using program slicing to generate targeted tests, followed by mutation-driven adversarial testing that synthesizes plausible-but-incorrect patches to expose semantic blind spots. 
% On SWE-Bench Verified (500 instances), SWE-ABS strengthens 50.2\% of instances, a $25.1\times$ improvement over prior work, and rejects 19.78\% of previously passing patches, causing the top-performing agent's success rate to drop from 78.80\% to 62.20\%.
% Further evaluation on a 150-instance multilingual subset of SWE-Bench Pro achieves a 64.7\% strengthening rate despite the top code agent scoring 33 percentage points lower on SWE-Bench Pro than on SWE-Bench Verified, revealing that task difficulty and test strength are orthogonal.

% exciting version-2

The SWE-Bench Verified leaderboard is approaching saturation, with the top system achieving 78.80\%.
However, we reveal that this performance is inflated: our re-evaluation demonstrates that one in five ``solved'' patches from the top-30 agents are semantically incorrect, passing only because weak test suites fail to expose their errors.
We present SWE-ABS, an adversarial framework that strengthens test suites through a two-stage pipeline:
(1) coverage-driven augmentation utilizing program slicing to target untested code regions, and (2) mutation-driven adversarial testing that synthesizes plausible-but-incorrect patches to expose semantic blind spots.
On SWE-Bench Verified (500 instances), SWE-ABS strengthens 50.2\% of instances (a $25.1\times$ improvement over prior work) and rejects 19.78\% of previously passing patches. Consequently, the top agent's score decreases from 78.80\% to 62.20\%, causing significant leaderboard reshuffling (e.g., the top-ranked agent drops to 5th place).

% Further evaluation on the more challenging SWE-Bench Pro (150-instance multilingual subset) achieves a 64.7\% strengthening rate despite the top agent scoring 33 percentage points lower than on SWE-Bench Verified, revealing that task difficulty and test strength are orthogonal.

\end{abstract}

\section{Introduction}
\label{sec:intro}

%  modified_figure  motivation
% Motivating Example Figure for Introduction
% Real SWE-Bench case: scikit-learn__scikit-learn-13241
\usetikzlibrary{positioning}

\begin{figure*}[t]
\centering
\footnotesize
\setlength{\fboxsep}{0pt}
\setlength{\fboxrule}{0.8pt}
\newlength{\codewidth}
\setlength{\codewidth}{5.2cm}

\begin{tikzpicture}[
    every node/.style={inner sep=0pt, outer sep=0pt}
]

% Define column positions
\def\colA{0}
\def\colB{5.8}
\def\colC{11.6}
\def\boxwidth{5.5}
\def\boxheight{5.5}
% ==================== COLUMN 1: Issue & Gold Patch ====================
\begin{scope}[shift={(\colA, 0)}]
    % Frame
    \fill[blue!3] (0,0) rectangle (\boxwidth, -\boxheight);
    \draw[blue!40, line width=0.8pt, rounded corners=3pt] (0,0) rectangle (\boxwidth, -\boxheight);

    % Header
    \fill[blue!15] (0,0) rectangle (\boxwidth, -0.6);
    \node[font=\footnotesize\bfseries] at (\boxwidth/2, -0.3) {(a) Issue \& Gold Patch};

    % Content
    \node[anchor=north west] at (0.2, -0.8) {
        \begin{minipage}{4.9cm}
        \scriptsize
        \textbf{Instance:} django\_\_django-10973\\[2pt]
        \textbf{Issue:} Refactor the PostgreSQL client backend to use subprocess.run with the PGPASSWORD environment variable
        \end{minipage}
    };

    % Gold patch label
    \node[anchor=north west, font=\scriptsize\bfseries, text=blue!70!black] at (0.2, -2.2) {
        Gold Patch \textcolor{green!60!black}{\ding{51}}
    };

    % Code block
    \node[anchor=north west] at (0.2, -2.5) {
        \begin{minipage}{4.9cm}
        \colorbox{gray!8}{%
        \begin{minipage}{\codewidth}
        \ttfamily\tiny
        subprocess\_env = os.environ.copy()\\
        if passwd:\\
        \hspace*{2mm}subprocess\_env['PGPASSWORD']=\textcolor{blue!70!black}{str(passwd)}\\
        ...\\
        subprocess.run(args, env=subprocess\_env)
        \end{minipage}}
        \end{minipage}
    };

    % Explanation
    \node[anchor=north west, font=\scriptsize, text=gray!60] at (0.2, -4.0) {
        \begin{minipage}{4.9cm}
        \textit{Correct:Explicit \texttt{str()} conversion ensures all environment variables are strings, satisfying \texttt{subprocess.run}'s requirement.}
        \end{minipage}
    };

\end{scope}

% ==================== COLUMN 2: Original Test ====================
\begin{scope}[shift={(\colB, 0)}]
    % Frame
    \fill[orange!3] (0,0) rectangle (\boxwidth, -\boxheight);
    \draw[orange!40, line width=0.8pt, rounded corners=3pt] (0,0) rectangle (\boxwidth, -\boxheight);

    % Header
    \fill[orange!15] (0,0) rectangle (\boxwidth, -0.6);
    \node[font=\footnotesize\bfseries] at (\boxwidth/2, -0.3) {(b) Original Test \& Agent Patch};

    % Test label
    \node[anchor=north west, font=\scriptsize\bfseries, text=orange!70!black] at (0.2, -0.8) {
        Original Test Suite
    };

    % Code block
    \node[anchor=north west] at (0.2, -1.2) {
        \begin{minipage}{4.9cm}
        \colorbox{orange!8}{%
        \begin{minipage}{\codewidth}
        \ttfamily\tiny
        def test\_basic():\\
        \hspace*{2mm}args, pgpassword =run(\\
        \hspace*{6mm}\{'password':\textcolor{red!70!black}{'secret'}\}\\
        \hspace*{2mm})\\
        \hspace*{2mm}assert pgpassword == 'secret'
        \end{minipage}}
        \end{minipage}
    };

    % Problem annotation
    \node[anchor=north west, font=\scriptsize, text=red!70!black] at (0.2, -2.6) {
        \begin{minipage}{4.9cm}
        \ding{55} \textbf{Semantic blind spot:}\\[2pt]
        Only string passwords are tested. Non-string inputs (e.g., integers) are never exercised.
        \end{minipage}
    };
        % Agent patch example
    \node[anchor=north west, font=\scriptsize] at (0.2, -3.6) {
        \begin{minipage}{4.9cm}
        \textbf{Agent Patch Example (TRAE + Doubao-Seed-Code)}\\[2pt]
        \colorbox{gray!8}{%
        \begin{minipage}{\codewidth}
        \ttfamily\tiny
        env = os.environ.copy()\\
        if passwd:\\
        \hspace*{2mm}env['PGPASSWORD'] = \textcolor{red!70!black}{passwd}
        \end{minipage}}
        \end{minipage}
    };

    % Status badge
    \node[anchor=south, font=\tiny\bfseries] at (\boxwidth/2, -5.3) {
        \colorbox{green!20}{\textcolor{black}{25 Wrong Patches PASS \textcolor{green!60!black}{\ding{51}}}}
    };
\end{scope}

% ==================== COLUMN 3: Strengthened Test ====================
\begin{scope}[shift={(\colC, 0)}]
    % Frame
    \fill[green!3] (0,0) rectangle (\boxwidth, -\boxheight);
    \draw[green!40, line width=0.8pt, rounded corners=3pt] (0,0) rectangle (\boxwidth, -\boxheight);

    % Header
    \fill[green!15] (0,0) rectangle (\boxwidth, -0.6);
    \node[font=\footnotesize\bfseries] at (\boxwidth/2, -0.3) {(c) Augmented Test};

    % Test label
    \node[anchor=north west, font=\scriptsize\bfseries, text=green!60!black] at (0.2, -0.8) {
        After Test Enhancement
    };

    % Code block
    \node[anchor=north west] at (0.2, -1.5) {
        \begin{minipage}{4.9cm}
        \colorbox{green!8}{%
        \begin{minipage}{\codewidth}
        \ttfamily\tiny
        def test\_password\_non\_string():\\
        \hspace*{2mm}\textcolor{green!60!black}{\# Adversarial: integer}\\
        \hspace*{2mm}args, pgpassword =run(\\
        \hspace*{6mm}\{'password':\textcolor{red!70!black}{123456}\}\\
        \hspace*{2mm})\\
        \hspace*{2mm}assert pgpassword ==\textcolor{green!60!black}{'123456'}
        \end{minipage}}
        \end{minipage}
    };

    % Impact annotation
    \node[anchor=north west, font=\scriptsize, text=green!60!black] at (0.2, -3.2) {
        \begin{minipage}{4.9cm}
        \textbf{Impact:}\\[2pt]
        Exposes missing type conversion logic in environment variable handling.
        \end{minipage}
    };

    % Status badge
    \node[anchor=south, font=\tiny\bfseries] at (\boxwidth/2, -5.3) {
        \colorbox{red!20}{\textcolor{black}{25 Wrong Patches FAIL \textcolor{red!60!black}{\ding{55}}}}
    };
\end{scope}

\end{tikzpicture}

\caption{
    \textbf{Motivating example: Weak tests enable silent failures.}
    (a)~A real Django issue (\texttt{django\_\_django-10973}) refactoring requires converting passwords to strings before passing them to \texttt{subprocess.run} via the \texttt{PGPASSWORD} environment variable, because \texttt{subprocess.run} only accepts string-valued environment variables.
    (b)~The original test suite tests only string passwords, enabling semantically incorrect agent patches to pass despite failing to handle non-string inputs.
    (c)~A strengthened test introduces non-string inputs, exposing the missing type conversion and correctly failing the invalid patches.
}

\label{fig:motivating}
\end{figure*}
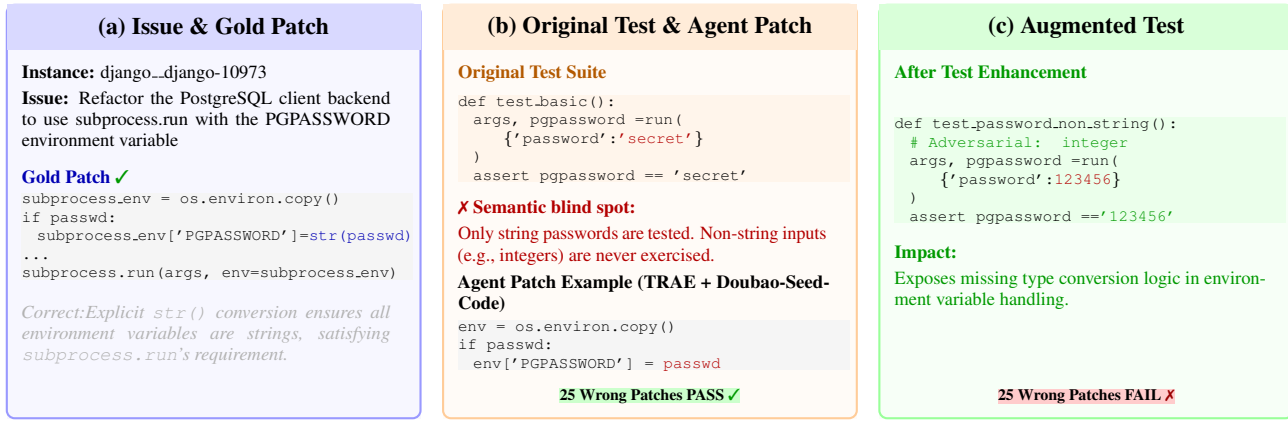

Reliable evaluation is fundamental to measuring progress in machine learning.
As benchmarks become the primary mechanism for comparing system capabilities, ensuring they possess sufficient \emph{discriminative power} (i.e., the ability to reject incorrect solutions while accepting correct ones) is critical.
Yet we reveal a systematic evaluation crisis: test suites in widely used benchmarks, often derived from real-world repositories, carry over the limitations of their original test designs, allowing semantically incorrect solutions to pass and inflating success rates.

We demonstrate this crisis in SWE-Bench~\cite{swebench}, a widely adopted benchmark for evaluating LLM-based software engineering agents.
The top-ranked code agent achieves a 78.80\% success rate, suggesting that the benchmark is approaching saturation.
Such apparent saturation is not unique to SWE-Bench: recent evaluation research shows that binary test-based protocols can mask substantial quality gaps, as models may optimize for merely test-passing behavior under weak discriminative signals~\cite{frontiercs}.
However, our analysis shows that 19.78\% of cases labeled as ``solved'' among the top-30 leaderboard agents are semantically incorrect.
Across 11,041 patches produced by the top-30 leaderboard agents that pass the original SWE-Bench Verified tests, we reject 2,184 (19.78\%) when applying strengthened test suites. This reduces the top agent's success rate by 16.6 percentage points, from 78.80\% to 62.20\%, causing it to drop from 1st to 5th place.

This crisis stems from a fundamental mismatch in objectives: SWE-Bench tests originate from development pull requests (PRs), which are designed to \emph{verify if a specific patch passes the predefined test suites} rather than to \emph{discriminate between all potential correct and incorrect solutions}.
This creates two systematic weaknesses: \emph{coverage gaps}, where tests miss patch-affected code entirely, and \emph{semantic blind spots}, where tests accept superficially correct behavior without verifying deeper semantic requirements.
Coverage gaps are straightforward to identify via static analysis; semantic blind spots, however, are more subtle and harder to detect.

To illustrate these weaknesses concretely, Figure~\ref{fig:motivating} presents a real example from the SWE-Bench dataset where the original test suite fails to expose a semantic blind spot.
The issue requires the PostgreSQL client backend to invoke \texttt{subprocess.run} for password retrieval, a function that requires a string argument.
However, the original test only validates string passwords, creating a semantic blind spot: 25 patches that omit the \texttt{str()} conversion successfully pass the original tests but fail our augmented test, which exercises non-string inputs.
As shown in Figure~\ref{fig:motivating}(b), a patch generated by \texttt{TRAE + Doubao-Seed-Code}~\cite{trae_agent, seed_coder}, the current top-performing agent on SWE-Bench, is plausible but incorrect: it addresses the refactoring request yet misses the string-type constraint, thereby passing the original tests while violating the actual requirement.

\paragraph{Adversarial Benchmark Strengthening.}
Prior work, such as UTBoost~\cite{utboost}, also attempts test augmentation by generating additional unit tests via LLMs, but strengthens only 10 out of 500 (2\%) instances on SWE-Bench Verified.
To uncover more vulnerabilities in the SWE-Bench test suite, we propose SWE-ABS, an \emph{adversarial} framework that actively attacks test suites to expose latent weaknesses, and then fortifies them.
Our two-stage approach combines complementary signals:

\emph{Stage~I: Coverage-Driven Augmentation} leverages program slicing to identify patch-affected code regions and generates tests to exercise them.
A \emph{test decoupling} mechanism is adopted to prevent generated tests from overfitting to the specific implementation details of the gold patch.

\emph{Stage~II: Mutation-Driven Adversarial Strengthening} targets subtle semantic blind spots. We synthesize ``mutant" patches that are plausible (i.e., they pass all existing tests) yet semantically incorrect.
By identifying mutants that evade existing tests, we generate targeted adversarial tests to reject them, effectively mirroring red-team/blue-team dynamics used in security testing.

\paragraph{Results.}
We evaluate SWE-ABS on SWE-Bench Verified~\cite{swebench_verified} (500 instances) and a subset of SWE-Bench Pro~\cite{swebench_pro} (150 instances). On SWE-Bench Verified, SWE-ABS strengthens 50.2\% of instances ($25.1\times$ improvement over UTBoost), and induces an average decline of 14.56 percentage points in resolve rates across systems, resulting in 30 rank changes within the top-30 leaderboard.
Cross-benchmark evaluation reveals a notable observation: although current code agents achieve substantially lower resolve rates on SWE-Bench Pro than on SWE-Bench Verified (45.89\% vs.\ 78.80\% for the top system), SWE-ABS attains a comparable strengthening rate on both benchmarks. This indicates that benchmarks that are more challenging for models do not necessarily provide higher-quality or more discriminative test cases.

\paragraph{Contributions.}
\begin{itemize}
    \item \textbf{SWE-ABS}, a two-stage adversarial benchmark strengthening framework that combines coverage-driven augmentation with mutation-driven adversarial testing. SWE-ABS strengthens 50.2\% of SWE-Bench Verified instances, a $25.1\times$ improvement over prior work (UTBoost).

    \item \textbf{Empirical evidence} that PR-driven test suites systematically lack discriminative power: our re-evaluation rejects 19.78\% of previously accepted solutions (2,184 out of 11,041 patches) and induces 30 rank changes among the top-30 leaderboard agents.

    \item \textbf{A counterintuitive finding} that task difficulty and test discriminativeness are orthogonal: SWE-ABS achieves comparable strengthening rates on both SWE-Bench Verified and the more challenging SWE-Bench Pro.

    \item \textbf{Strengthened test suites} for SWE-Bench Verified (500 instances) and a subset of SWE-Bench Pro (150 instances), included in supplementary materials to facilitate more rigorous evaluation of future code agents.
\end{itemize}

\section{Related Work}

\begin{figure*}[t]
    \centering
    \includegraphics[
        width=0.9\textwidth,
        trim=5mm 5mm 5mm 5mm,
        % clip
    ]{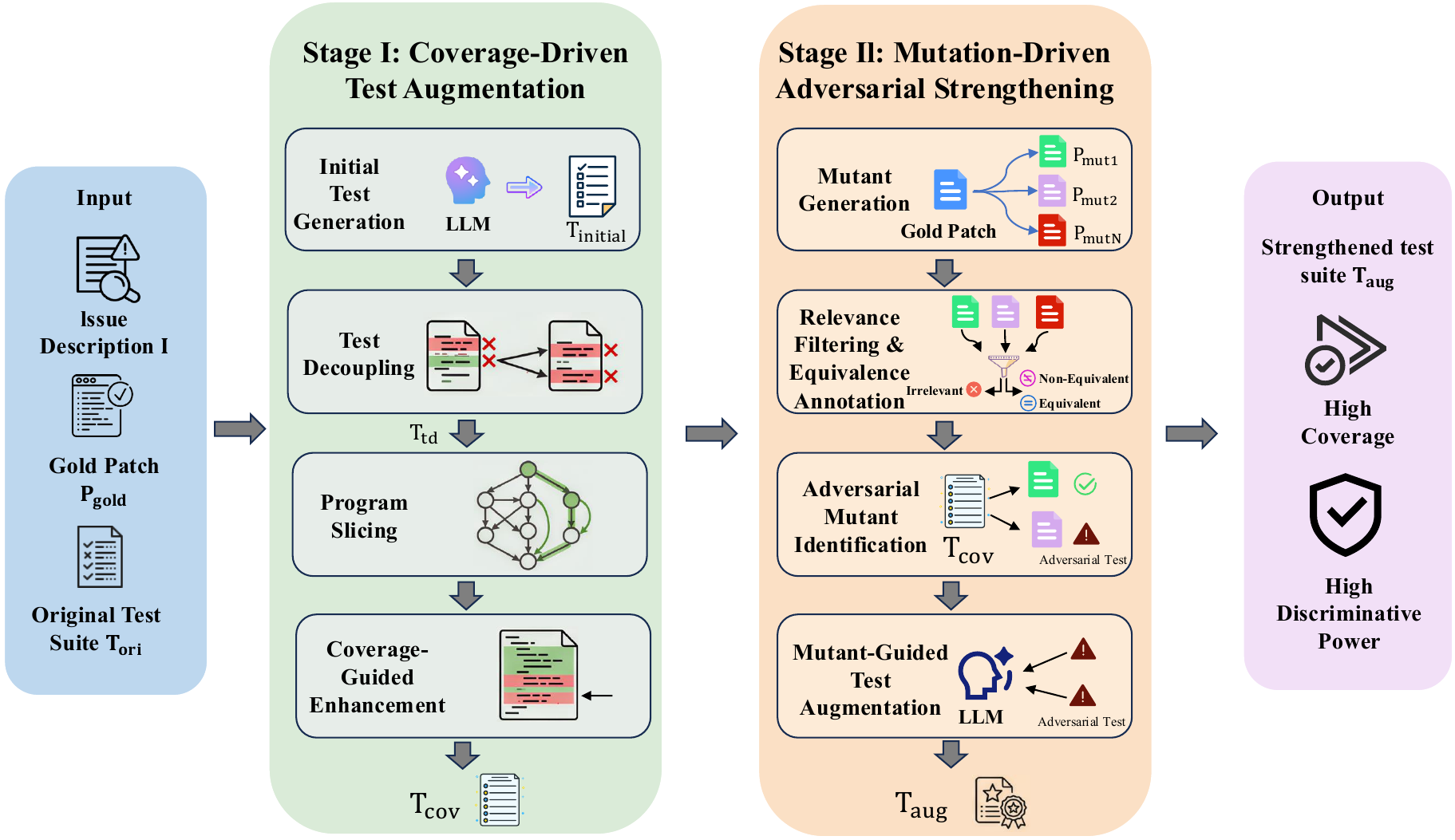}
    \caption{Overview of the SWE-ABS framework.}
    \label{fig:overview}
\end{figure*}

\paragraph{Benchmarks for Code Generation.}
Code generation evaluation has evolved from function-level benchmarks like HumanEval~\cite{humaneval} and MBPP~\cite{mbpp_program_synthesis} to repository-level challenges.
SWE-Bench~\cite{swebench} and its variants have progressively improved test validity (SWE-Bench Verified~\cite{swebench_verified}), language coverage (Multi-SWE-Bench~\cite{multi_swebench}), contamination resistance (SWE-Bench Live~\cite{swebench_live}), and task difficulty (SWE-Bench Pro~\cite{swebench_pro}).
SWE-Bench Pro curates more challenging instances from strong copyleft licenses codebases spanning Python, JavaScript, TypeScript, and Go, requiring larger code modifications (averaging 107 lines across 4 files).
However, all SWE-Bench variants rely on PR-originated test suites designed to verify specific patches rather than discriminate among alternative solutions, leaving coverage gaps and semantic blind spots that allow incorrect patches to pass.
We select SWE-Bench Verified and SWE-Bench Pro as evaluation benchmarks to assess our method across both established and multilingual settings.

\paragraph{Test Augmentation.}
Test-based evaluation of program repair faces a fundamental challenge: patches that are plausible (passing all tests) may not be correct (genuinely fixing the bug), as they can overfit to insufficient test suites~\cite{patch_plausibility_and_correctness, genprog_overfitting}.
EvalPlus~\cite{eval_plus} demonstrated test insufficiency in established benchmarks, augmenting HumanEval~\cite{humaneval} and MBPP~\cite{mbpp_program_synthesis} with 80$\times$ more tests.
UTBoost~\cite{utboost} pioneered test augmentation for SWE-Bench, exposing 345 previously undetected incorrect patches on SWE-Bench.
However, these approaches generate tests without explicitly analyzing patch-relevant code regions or targeting semantic correctness, limiting their ability to systematically expose both coverage gaps and semantic blind spots.

\paragraph{Mutation Testing.}
Mutation testing~\cite{mutation_testing,mutation_testing_survey} evaluates test quality by introducing artificial faults, traditionally using syntactic operators (e.g., replacing \texttt{+} with \texttt{-}).
Recent work leverages LLMs to generate more realistic mutants for the evaluation of test adequacy~\cite{mutation_testing_study_llm} and the generation of targeted tests~\cite{mutation_guided_llm_based_test_generation_at_meta}.
SWE-ABS extends this paradigm by synthesizing LLM-based semantic mutations that represent plausible but incorrect patches that pass existing test suites but violate actual requirements, then generating tests that expose these blind spots.

\section{Method}
\label{sec:method}

SWE-Bench relies on developer-written tests that provide necessary but insufficient conditions for patch correctness: a correct patch must pass these tests, but passing does not guarantee the correctness.
As discussed in Section~\ref{sec:intro}, this leads to two systematic weaknesses: \emph{coverage gaps}, where tests fail to exercise patch-affected code regions, and \emph{semantic blind spots}, where tests accept superficially correct behavior without verifying deeper semantic requirements.

We address these complementary weaknesses through a two-stage adversarial strengthening framework (Figure~\ref{fig:overview}).
Stage~I targets coverage gaps by using program slicing to identify patch-relevant code and generating tests that exercise those regions, while decoupling these tests from gold-patch-specific behaviors to avoid overfitting.
Stage~II targets semantic blind spots by synthesizing plausible but incorrect mutant patches that evade existing tests, then generating adversarial tests to reject them.

Each SWE-Bench instance consists of an issue description $I$, a gold patch $P_{\text{gold}}$, and an original test suite $T_{\text{ori}} = T_{\text{base}} \cup T_{\text{patch}}$, where $T_{\text{base}}$ denotes the repository's existing tests and $T_{\text{patch}}$ denotes the tests added in the pull request.
Our goal is to construct an augmented test suite $T_{\text{aug}}$ that rejects incorrect patches while accepting correct alternative implementations.

% \paragraph{Notation.}
% Throughout this section, we use the following notation: $T_{\text{initial}}$ denotes tests from initial generation, $T_{\text{td}}$ denotes tests after decoupling, $L_{\text{rel}}$ denotes patch-relevant code lines, $T_{\text{cov}}$ denotes the coverage-enhanced test suite from Stage~I, $P_{\text{mut}}$ denotes generated mutant patches, $P_{\text{rel}}$ denotes filtered relevant mutants, $P_{\text{adv}}$ denotes adversarial mutants, and $T_{\text{aug}}$ denotes the final augmented test suite.

% We denote the original SWE-Bench test suite as $T_{\text{ori}}$, the coverage-enhanced suite from Stage~I as $T_{\text{cov}}$, and the final augmented suite as $T_{\text{aug}}$.
% Other notations are introduced where they first appear.

\subsection{Stage~I: Coverage-Driven Test Augmentation}
\label{subsec:coverage}

Coverage gaps arise when tests fail to exercise patch-affected code regions, allowing incorrect patches to pass undetected.
This stage addresses such gaps through four steps: (1) initial test generation creates tests targeting the issue, (2) test decoupling removes gold-patch-specific implementation details, (3) program slicing identifies patch-relevant code regions $L_{\text{rel}}$, and (4) coverage-guided enhancement adds tests for uncovered lines in $L_{\text{rel}}$.

\subsubsection{Generating Initial Tests}
\label{subsec:initial-gen}

In the first step, we prompt the LLM with the issue $I$, gold patch $P_{\text{gold}}$, and patch tests $T_{\text{patch}}$ to synthesize the initial augmented test suite $T_{\text{initial}}$ (all prompt templates can be found in Appendix~\ref{app:prompts}).
The model is instructed to generate diverse test cases that cover corner cases.
Providing $T_{\text{patch}}$ as context helps it understand the intended test style.

% In the first step, we prompt the LLM to generate various test cases and cover the corner cases.
% Using the issue $I$, gold patch $P_{\text{gold}}$, and patch tests $T_{\text{patch}}$ as input, the model synthesizes the initial augmented test suite $T_{\text{initial}}$.
% Providing $T_{\text{patch}}$ as context helps the LLM understand the test's intent and style.
% The agent generates tests targeting edge cases, boundary conditions, and interaction scenarios beyond $T_{\text{init}}$.

\subsubsection{Decoupling Gold-Patch Dependencies}
\label{subsec:test-decouple}
% In SWE-Bench, some test cases are overly specific to the gold patch.
While initial test generation produces diverse tests, some may be overly specific to the gold patch (An example in Appendix~\ref{app:test_decoupling_cases}).
To mitigate this issue in $T_{\text{initial}}$, we introduce a test decoupling module.
We first prompt the LLM to detect the tests that are over-specialized to gold patch, such as hard-coded error messages or implementation-specific side effects, which may reject some valid alternative fixes.
These problematic tests are then refined by prompting the LLM to generalize them, ensuring that their core purpose is to verify whether the original issue has been resolved, rather than enforcing a particular implementation.
After test decoupling, we verify whether the gold patch still passes the refined tests.
We define the subset of refined tests that the gold patch passes as $T_{\text{td}}$.

\subsubsection{Program Slicing}
\label{subsec:code-region}
To target test generation toward patch-relevant code, we use intraprocedural program slicing (details in Appendix~\ref{app:slicing_implementation}).
We construct a program dependence graph that captures data and control dependencies, then compute patch-relevant lines $L_{\text{rel}}$ as the forward and backward slice from modified lines, capturing both upstream dependencies and downstream dependents.
To balance precision and scalability, we restrict analysis to statements within the same function or class as the patch, avoiding prohibitively expensive whole-program analysis.

\subsubsection{Coverage-Guided Augmentation}
\label{subsec:coverage-enhance}
With patch-relevant lines identified, we now enhance coverage of these regions.
For each test $t \in T_{\text{td}}$, we measure its coverage over the patch-relevant lines $L_{\text{rel}}$ by using a code coverage tool, where the details are in Appendix~\ref{app:coverage_calculation}.
We feed the identified uncovered lines to the LLM and prompt it to generate tests that address the uncovered code branches.
We define the coverage-guided enhanced test as follows:
\[
T_{\text{cov}} = T_{\text{td}} \;\cup\;
\mathrm{LLM}_{\text{cov}}\Big( L_{\text{rel}} \setminus \bigcup_{t \in T_{\text{td}}} \mathrm{Cov}(t) \Big),
\]
where $\mathrm{Cov}(t) \subseteq L_{\text{rel}}$ denotes the set of lines exercised by test $t$, and $\mathrm{LLM}_{\text{cov}}$ generates tests to cover the specified lines.
This coverage-guided enhancement aims to maximize coverage of patch-relevant lines.

\subsection{Stage~II: Mutation-Driven Adversarial Strengthening}
\label{subsec:mutation}

Stage~I produces $T_{\text{cov}}$, which ensures \emph{reachability} of patch-relevant code.
However, reachability alone does not guarantee \emph{observability} of semantic faults: a faulty implementation may execute the same control-flow paths as the correct patch yet produce incorrect states or outputs that existing assertions fail to detect.
To address this limitation, Stage~II employs mutation-driven adversarial strengthening to target \emph{semantic blind spots}, i.e., scenarios where incorrect implementations satisfy coverage requirements but fail to correctly resolve the issue.

\subsubsection{Mutant Generation}
\label{subsec:mutant-gen}

We employ an LLM-based mutation generator, denoted as $\mathrm{LLM}_{\text{mut}}$, to produce mutant patches $P_{\text{mut}}$.
Given a gold patch $P_{\text{gold}}$, $\mathrm{LLM}_{\text{mut}}$ generates mutants that introduce subtle semantic faults while passing the original test suite $T_{\text{ori}}$:
\[
P_{\text{mut}} = \mathrm{LLM}_{\text{mut}}(P_{\text{gold}}).
\]

% These mutants mimic plausible but incorrect fixes that satisfy $T_{\text{init}}$, serving as adversarial probes to identify weaknesses not addressed by Stage I's coverage augmentation.

\subsubsection{Relevance Filtering and Equivalence Annotation}
\label{subsec:mutant-filter}
To mitigate hallucinated or irrelevant mutations, we apply an LLM-based filtering module that evaluates each mutant along two orthogonal dimensions:
(1) \emph{issue relevance}, i.e., whether the mutation targets behaviors or code regions implicated by the original issue, and
(2) \emph{semantic equivalence}, i.e., whether the mutation is functionally equivalent to the gold patch.
% Each judgment is independently performed by k LLMs (we set k as 3), and the final decision is determined via majority voting.
For each dimension, we obtain independent judgments from $k$ LLMs (with $k=3$ in our experiments) and apply majority voting to determine the final label.
A comparison with human annotations, along with illustrative examples, is provided in the Appendix~\ref{app:mutation_case}.
We first filter out issue-irrelevant mutants to obtain a subset of issue-relevant mutants $P_{\text{rel}} \subseteq P_{\text{mut}}$.
Then, we label the semantic equivalence of each patch $p$, where $\mathrm{equiv}(p)$ denotes semantic equivalence and $\neg\mathrm{equiv}(p)$ denotes semantic non-equivalence.
% Each $p \in P_{\text{rel}}$ is then annotated with an equivalence predicate $\mathrm{equ}(p) \in \{\text{true}, \text{false}\}$ for downstream categorization and evaluation.

\subsubsection{Identifying Adversarial Mutants}
\label{subsec:adversarial-id}

After obtaining the filtered and annotated mutants $P_{\text{rel}}$, we identify which mutants expose weaknesses in the current test suite $T_{\text{cov}}$ by checking whether each mutant $p$ passes or fails the tests.
An ideal test suite should accept semantically equivalent mutants and reject non-equivalent ones.
Discrepancies between expected and actual outcomes reveal test suite weaknesses:
\begin{itemize}
    \item \textbf{False negatives}: Equivalent mutants (correct alternatives) that fail $T_{\text{cov}}$, indicating tests are overly specific to the gold patch.
    \item \textbf{False positives}: Non-equivalent mutants (incorrect patches) that pass $T_{\text{cov}}$, revealing semantic blind spots.
\end{itemize}

We partition these discrepancy-inducing mutants into two categories based on their failure mode:
\begin{align}
P_{\text{adv}}^{\text{FN}} &= \left\{ p \in P_{\text{rel}} \mid \mathrm{equiv}(p) \wedge p \not\models T_{\text{cov}} \right\}, \\
P_{\text{adv}}^{\text{FP}} &= \left\{ p \in P_{\text{rel}} \mid \neg\mathrm{equiv}(p) \wedge p \models T_{\text{cov}} \right\}.
\end{align}
For each false-negative mutant $p \in P_{\text{adv}}^{\text{FN}}$, we further identify the specific tests responsible for the incorrect rejection:
\[
T_{\text{fn}} = \bigcup_{p \in P_{\text{adv}}^{\text{FN}}} \left\{ t \in T_{\text{cov}} \mid p \not\models t \right\}.
\]

\subsubsection{Mutant-Guided Test Augmentation}
\label{subsec:mutant-augment}

\paragraph{Fixing False Negatives.}
For tests in $T_{\text{fn}}$ that incorrectly reject equivalent mutants, we invoke an LLM-based test fixer $\mathrm{LLM}_{\text{fix}}$ to generalize them:
\[
T_{\text{fn}}' = \mathrm{LLM}_{\text{fix}}\left( T_{\text{fn}}, P_{\text{adv}}^{\text{FN}} \right).
\]
The fixer relaxes overly specific assertions while preserving the tests' ability to verify the original issue resolution.

\paragraph{Fixing False Positives.}
For each false-positive mutant $p \in P_{\text{adv}}^{\text{FP}}$, we invoke an LLM-based test generator $\mathrm{LLM}_{\text{aug}}$ to produce targeted tests that reject $p$:
\[
T_{\text{mut}}(p) = \mathrm{LLM}_{\text{aug}}(p, T_{\text{cov}}).
\]

\paragraph{Final Test Suite.}
We construct the final augmented test suite by replacing the overly strict tests with their generalized versions and adding the new discriminative tests:
\[
T_{\text{aug}} = \left( T_{\text{cov}} \setminus T_{\text{fn}} \right) \cup T_{\text{fn}}' \cup \bigcup_{p \in P_{\text{adv}}^{\text{FP}}} T_{\text{mut}}(p).
\]
This adversarial augmentation complements the coverage-driven tests from Stage I, yielding a test suite that is both comprehensive in coverage and robust against plausible but incorrect fixes.

\section{Experiments}
\label{sec:experiments}

Our experiments aim to answer the following research questions:

\begin{itemize}
    \item \textbf{RQ1 (Effectiveness):} Does SWE-ABS significantly improve the discriminative power of SWE-Bench test suites in distinguishing correct fixes from incorrect patches?
    \item \textbf{RQ2 (Generalization):} Does SWE-ABS generalize across different benchmarks and underlying base models?
    % \item \textbf{RQ3 (Quality Analysis):} How robust are SWE-ABS tests to alternative correct implementations while detecting semantic errors?
    \item \textbf{RQ3 (Test Quality):} Do SWE-ABS-augmented tests maintain robustness to alternative correct implementations while effectively detecting semantic errors?
    \item \textbf{RQ4 (Ablation Study):} How do the coverage-driven and mutation-driven stages individually and jointly contribute to effectiveness?
    % \item \textbf{RQ3 (Quality Analysis):} Are SWE-ABS-augmented tests robust to alternative correct implementations? What types of semantic errors does SWE-ABS detect that are missed by original tests?
    % \item \textbf{RQ4 (Ablation Study):} How do the coverage-driven and mutation-driven stages individually contribute to strengthening effectiveness? Are they complementary?
\end{itemize}

\begin{table*}[t]
\caption{Agent-level results on SWE-Bench Verified (Top-5 systems) for TRAE~\cite{trae_agent} with Doubao-Seed-Code~\cite{seed_coder}, Live-SWE-agent~\cite{live_swe_agent} with Gemini~\cite{gemini} 3 Pro Preview, 2025-11-18, Atlassian Rovo Dev~\cite{atlassian_rovo_dev} (2025-09-02), EPAM AI/Run Developer Agent~\cite{epam_ai_run_developer_agent} (v20250719) with Claude 4 Sonnet and ACoder~\cite{acoder}.
Scores denote resolve rates (\%) under the \emph{original} (SWE-Bench Verified), UTBoost, and SWE-ABS test suites.
Drop denotes the per-agent decrease in resolve rate (in percentage points).
Rank indicates changes in leaderboard position among all 30 agents.
Full results are in Appendix~\ref{app:full_performance}.
}
\centering
% \footnotesize
\resizebox{1.0\linewidth}{!}{
\begin{tabular}{ l c ccc ccc }
\toprule
\multirow{2.5}{*}{\textbf{Agent}}
&
\textbf{Original}
& \multicolumn{3}{c}{\textbf{UTBoost}}
& \multicolumn{3}{c}{\textbf{SWE-ABS (Ours)}} \\
\cmidrule(lr){2-2} \cmidrule(lr){3-5} \cmidrule(lr){6-8}
& \textbf{Score}
& \textbf{Score} & \textbf{Drop} & \textbf{Rank}
& \textbf{Score} & \textbf{Drop} & \textbf{Rank} \\
\midrule
TRAE + Doubao-Seed-Code
& 78.80
& 77.80 & 1.00 & 1$\to$1
& 62.20 & 16.60 & 1$\to$5 \\
live-SWE-agent + Gemini 3 Pro Preview (2025-11-18)
& 77.40
& 76.80 & 0.60 & 2$\to$2
& 65.40 & 12.00 & 2$\to$1 \\
Atlassian Rovo Dev (2025-09-02)
& 76.80
& 75.60 & 1.20 & 3$\to$5
& 61.20 & 15.60 & 3$\to$8 \\
EPAM AI/Run Developer Agent v20250719 + Claude 4 Sonnet
& 76.80
& 76.20 & 0.60 & 4$\to$3
& 63.80 & 13.00 & 4$\to$3 \\
ACoder
& 76.40
& 75.80 & 0.60 & 5$\to$4
& 61.80 & 14.60 & 5$\to$6 \\

\bottomrule
\end{tabular}
}% resize
\label{tab:rq1_system}
\end{table*}

\begin{table*}[t]
\caption{Benchmark-level results on SWE-Bench Verified.
Str.\ (Strengthened) denotes the number of instances where at least one previously passing patch fails under augmented tests.
Avg.\ Drop is measured in percentage points.
Patch Kill denotes the total number of patches rejected by augmented tests.
\# Rank Changes denotes the number of agents whose leaderboard position changed.
Spearman $\rho$ indicates the rank correlation coefficient between old and new leaderboard positions ($\rho = 1$ indicates perfect stability).
SWE-ABS* excludes the 53 instances with overfitting tests (Section~\ref{subsubsec:robustness});
SWE-ABS reports results after fixing the overfitting tests in those instances.}
\centering
\small
\begin{tabular}{lccccccc}
\toprule
\textbf{Augmentation} &
\textbf{Str.} &
\textbf{Avg. Drop} &
\textbf{Patch Kill} &
\textbf{\# Rank Changes} &
\textbf{Spearman $\rho$} \\
\midrule
UTBoost & 10 / 500  &  0.70   & 105/11,041  & 24 & 0.98\\
SWE-ABS* & 206 / 500 &  11.22  & 1683/11,041 & 25 & 0.86\\
SWE-ABS & 251 / 500 &  14.56  & 2184/11,041 & 30 & 0.82\\
\bottomrule
\end{tabular}

\label{tab:rq1_benchmark}
\end{table*}

\subsection{Experimental Setup}
\label{subsec:exp-setup}

\paragraph{Benchmarks and Evaluation Protocol.}
We evaluate on two benchmarks:
\textbf{SWE-Bench Verified}~\cite{swebench_verified}, containing 500 human-validated Python instances where the current best agent achieves 78.80\% resolve rate; and \textbf{SWE-Bench Pro}~\cite{swebench_pro}, a harder, contamination-resistant benchmark with 731 multi-language instances (Python, JavaScript, Go, TypeScript) from strong copyleft licenses repositories, where the best code agent achieves only 45.89\%.
From SWE-Bench Pro, we stratify by programming language and randomly sample 150 instances from those for which at least one leaderboard agent produces a patch that passes the original tests.
A per-language breakdown of the SWE-Bench Pro sampling distribution is provided in Table~\ref{tab:swe_pro_per_language} (Appendix~\ref{app:results}).
For each instance, we collect patches from leaderboard agents that pass the original tests (top-30 for SWE-Bench Verified, all 13 available for SWE-Bench Pro) and re-evaluate them under augmented tests.
Leaderboard submissions and patch data were collected as of December 31, 2025.
% RQ1 uses all 500 SWE-Bench Verified instances (11,041 total patches), while RQ2 samples 150 instances from SWE-Bench Pro.

\paragraph{Evaluation Metrics.}
We report four primary metrics:
(1)~\textbf{Strengthened (Str.)}: the number of instances where at least one previously passing patch fails under augmented tests;
(2)~\textbf{Drop}: the decrease in resolve rate (in percentage points);
(3)~\textbf{Patch Kill}: the total number of patches rejected by augmented tests;
(4)~\textbf{Spearman $\rho$}: the rank correlation coefficient between the original and augmented leaderboard rankings, measuring the extent of ranking stability ($\rho = 1$ indicates perfect stability, while lower values indicate greater reordering; detailed calculation in Appendix~\ref{app:spearman_calculation}).

\paragraph{Model Usage.}
We use GPT-5~\cite{gpt-5} as the default base model throughout all stages of SWE-ABS.
To demonstrate generalizability, we also conduct controlled experiments with an alternative open-source model, GLM-4.7~\cite{glm-47}.
Unless otherwise specified, all reported results are obtained using GPT-5.
% A detailed cost-efficiency analysis is provided in Appendix~\ref{app:cost}.

\paragraph{Patch-Relevant Code Identification.}
We perform intraprocedural program slicing using Tree-sitter~\cite{tree-sitter} to identify code regions that are data-flow or control-flow dependent on a given patch (details in Appendix~\ref{app:slicing_implementation}).

All hyperparameters used in SWE-ABS, including temperature and maximum number of tries, are provided in Appendix~\ref{app:setup}.
All key prompts used for SWE-ABS are provided in Appendix~\ref{app:prompts}.

\subsection{RQ1: Effectiveness on SWE-Bench Verified}
\label{subsec:rq1}

We compare SWE-ABS against UTBoost~\cite{utboost} on all 500 SWE-Bench Verified instances.

\paragraph{Results.}
As shown in Tables~\ref{tab:rq1_system} and~\ref{tab:rq1_benchmark}, SWE-ABS substantially outperforms UTBoost across all metrics.
At the benchmark level, SWE-ABS strengthens 251/500 instances (50.2\%) compared to only 10 (2\%) under UTBoost.
Moreover, it identifies and rejects 2,184 of the 11,041 previously passing patches (19.78\%) as insufficiently correct.
Detailed per-repository breakdowns are provided in Table~\ref{tab:per-repo} (Appendix~\ref{app:results}).
Table~\ref{tab:rq1_benchmark} also includes SWE-ABS*, which simply excludes the 53 overfitting instances identified in Section~\ref{subsubsec:robustness} without correction; SWE-ABS reports the final results after those overfitting tests are fixed.

At the agent level, SWE-ABS induces an average resolve-rate drop of 14.56 percentage points, compared to just 0.70 percentage points for UTBoost.
This drop triggers significant leaderboard reordering, exposing evaluation instability under weak test suites.
The Spearman rank correlation $\rho$ decreases from 0.98 (UTBoost) to 0.82 (SWE-ABS), indicating substantial ranking instability when test suites are properly strengthened.
Among the top-5 agents, four experience significant rank changes under SWE-ABS (Table~\ref{tab:rq1_system}).
Notably, TRAE~\cite{trae_agent} drops from 1st to 5th despite achieving the highest original score (78.80\%), suggesting that its patches, while strong in many cases, may fail to handle certain corner cases effectively.
In contrast, live-SWE-Agent~\cite{live_swe_agent} climbs from 2nd to 1st, indicating more robust and reliable patch generation overall.
This instability underscores that current leaderboard rankings may not reflect true agent capabilities.

The full Top-30 leaderboard changes, along with results for Bash-only agents on SWE-Bench Verified, are reported in Appendix~\ref{app:full_performance} and Appendix~\ref{app:Bash_Only_full_performance}, respectively.
A detailed cost analysis in Appendix~\ref{app:cost} shows that SWE-ABS achieves 16$\times$ better cost efficiency than UTBoost (\$4.98 vs.\ \$80.00 per strengthened instance).

\subsection{RQ2: Generalization}
\label{subsec:rq2}

\subsubsection{Cross-Benchmark Generalization}
\label{subsec:rq2-cross-benchmark}

To verify that SWE-ABS generalizes beyond Python-only SWE-Bench Verified, we evaluate on SWE-Bench Pro, a harder, contamination-resistant benchmark with multi-language instances (Python, JavaScript, Go, TypeScript).

\paragraph{Results.}
% Table~\ref{tab:rq2_cross_benchmark} shows that SWE-ABS consistently strengthens test suites across benchmarks with substantially different difficulty levels.
% Despite the top-system resolve rate on SWE-Bench Pro being 33 points lower than on SWE-Bench Verified (45.89\% vs.\ 78.80\%), SWE-ABS induces comparable strengthening effects: on the 150-instance SWE-Bench Pro subset, the top system drops from 70.00\% to 45.33\% (24.67\% drop), with an average drop of 16.46\% across all systems, closely matching the 14.56\% average drop on Verified.
Table~\ref{tab:rq2_cross_benchmark} (Appendix~\ref{app:generalization}) shows that SWE-ABS consistently strengthens test suites across benchmarks with substantially different difficulty levels.
Despite the top-system resolve rate on SWE-Bench Pro being 33 points lower than on SWE-Bench Verified (45.89\% vs.\ 78.80\%), SWE-ABS induces comparable strengthening effects: the average resolve-rate drop is 16.46 percentage points on the SWE-Bench Pro subset vs.\ 14.56 percentage points on SWE-Bench Verified, indicating that the discriminative power of augmented tests generalizes.
This validates two key insights:
(1)~\textbf{Task difficulty $\neq$ test strength}: a benchmark can be hard yet still have weak test suites;
(2)~\textbf{Method generalization}: SWE-ABS addresses fundamental test inadequacy that persists across benchmarks, not artifacts specific to Verified.
The consistent effectiveness on contamination-resistant SWE-Bench Pro instances further mitigates data leakage concerns.
We further report per-language breakdowns on SWE-Bench Pro in Table~\ref{tab:swe_pro_per_language} (Appendix~\ref{app:results})

\subsubsection{Base Model Generalization} \label{subsec:rq2-base-model}

To evaluate the generalizability of SWE-ABS beyond GPT-5, we conduct a controlled experiment with the open-source model GLM-4.7, keeping all non-model components identical.
SWE-ABS is applied to a random subset of 50 instances from SWE-Bench Verified under both model configurations.

\paragraph{Results.}
% As shown in Table~\ref{tab:base-model-generalization}, SWE-ABS exhibits comparable strengthening behavior across the two base models.
As shown in Table~\ref{tab:base-model-generalization} (Appendix~\ref{app:generalization}), SWE-ABS exhibits comparable strengthening behavior across the two base models.
Using GLM-4.7 yields an identical number of strengthened instances (31 vs.\ 31 out of 50) and a lower average drop in resolve rate (15.93 percentage points vs.\ 19.40 percentage points) compared to GPT-5.
These results suggest that the effectiveness of SWE-ABS is not tightly coupled to a specific proprietary base model and that the framework maintains its discriminative impact when instantiated with an alternative open-source model.

\subsection{RQ3: Test Quality} 
\label{subsec:rq3}

We analyze the quality of SWE-ABS-augmented tests along two complementary dimensions: \textbf{(1) robustness} to alternative correct implementations (avoiding false rejections), and \textbf{(2) semantic error coverage} (characterizing true rejections).

\subsubsection{Robustness to Alternative Correct Fixes}
\label{subsubsec:robustness}

A key concern is whether aggressive test-suite strengthening leads to overfitting to the gold patch, thereby inadvertently rejecting valid alternative fixes.
% Table~\ref{tab:test_admission} (Appendix~\ref{app:test_quality}) reports patch admission results under the original test suite ($T_{\text{ori}}$) and the strengthened suite ($T_{\text{aug}}$).
% Among the 463 instances that admit at least one patch under $T_{\text{ori}}$, 403 (87\%) continue to admit at least one patch under $T_{\text{aug}}$, while 60 instances reject all candidate patches.

% Manual inspection of these 60 instances revealed that 24 contain overfitting tests encoding gold-patch-specific behavior (see an example in Appendix~\ref{app:test_strengthening_failure}), while the remaining 36 reflect legitimate strengthening that correctly filter out spurious fixes.
% For these 24 overfitting cases, we disable test augmentation and revert to the original test suites; Tables~\ref{tab:rq1_system} and~\ref{tab:rq1_benchmark} report results after this correction.
% Overall, SWE-ABS achieves a conservative false-negative rate of 10.6\% (53/500).

Manual inspection of all 500 SWE-Bench Verified instances revealed that 53 contain overfitting tests encoding gold-patch-specific behavior (see an example in Appendix~\ref{app:test_strengthening_failure}), while the remaining 447 reflect legitimate strengthening that correctly filter out spurious fixes.
For these 53 overfitting cases, we correct the augmented tests to  revise or remove the gold-patch-specific assertions; Tables~\ref{tab:rq1_system} and~\ref{tab:rq1_benchmark} report results after this correction.
Overall, SWE-ABS achieves a conservative false-negative rate of 10.6\% (53/500).

\subsubsection{Semantic Error Coverage}

To understand what errors SWE-ABS detects, we manually analyzed 100 agent-generated patches rejected by $T_{\text{aug}}$ but accepted by $T_{\text{ori}}$.
Two authors independently categorized root causes, resolving disagreements through discussion.

% Table~\ref{tab:error_taxonomy} summarizes the distribution of error types; representative examples illustrating each category are provided in Appendix~\ref{subsubsec:agent_patch_failure}.
Table~\ref{tab:error_taxonomy} (Appendix~\ref{app:test_quality}) summarizes the distribution of error types; representative examples illustrating each category are provided in Appendix~\ref{subsubsec:agent_patch_failure}.
\textbf{Logic errors} (47\%) dominate, producing patches that pass the original test suite yet implement fundamentally incorrect logic, leading to failures on unseen scenarios.
\textbf{Incomplete fixes} (35\%) are also frequent, where patches address primary cases but overlook edge conditions, suggesting that AI systems tend to generate ``shallow'' solutions that satisfy explicit tests without achieving semantic completeness.
Type mismatches (7\%), boundary violations (6\%), off-by-one errors (1\%), and other errors (4\%) account for the remaining 18\% of failures.
% These findings reveal systematic weaknesses masked by inadequate test suites: current AI systems prioritize test-passing over semantic correctness, effectively ``teaching to the test.''

\paragraph{Implications.}
The dominance of logic errors (47\%) and incomplete fixes (35\%) reveals a systematic weakness in current AI code generation: agents tend to produce ``shallow'' solutions that satisfy explicit test cases without achieving semantic completeness.
This behavior aligns with training paradigms that optimize for test-passing reward signals, inadvertently encouraging agents to ``teach to the test'' rather than reason about program semantics.

The low prevalence of syntactic errors (type mismatches, boundary violations, and off-by-one errors account for 14\%) suggests that modern LLMs have largely solved surface-level code correctness.
The remaining challenge lies in \emph{semantic reasoning}, which means understanding what the code should do, not just what makes tests pass.
This finding has implications for future benchmark design: effective evaluation requires tests that probe semantic understanding rather than just syntactic correctness.

\subsection{RQ4: Ablation Study}
\label{subsec:rq4}

\begin{table}[t]
\caption{Ablation study on the effect of coverage-driven and mutation-based augmentations.
Str.\ (Strengthened) denotes the number of instances where at least one previously passing patch fails.
Initial denotes baseline tests generated without any augmentation stage.
Avg.\ Drop is measured in percentage points.}
\centering
\small
\begin{tabular}{lcc}
\toprule
\textbf{Setting} & \textbf{Str.} & \textbf{Avg.\ Drop} \\
\midrule
Initial             & 190 & 10.34 \\
Coverage         & 198 & 11.30 \\
\textbf{Coverage + Mutation} & \textbf{251} & \textbf{14.56} \\
\bottomrule
\end{tabular}

\label{tab:rq4-ablation-study}
\end{table}

% \paragraph{Coverage-Driven Augmentation (Stage~I).}

% \paragraph{Mutation-Driven Adversarial Strengthening (Stage~II).}

\paragraph{Coverage-Driven Augmentation (Stage~I).}
We first evaluate the effect of applying only the coverage-driven augmentation stage.
This stage aims to improve code reachability by identifying patch-relevant program regions via slicing and generating additional tests targeting previously uncovered code paths.
As shown in Table~\ref{tab:rq4-ablation-study}, this strategy yields a modest improvement on both reported metrics compared to the initial generated tests.
The number of strengthened instances increases from 190 to 198, while the average drop in resolve rate improves from 10.34 to 11.30 percentage points.
These results indicate that exercising patch-relevant but previously uncovered code regions helps reveal certain semantic faults that were masked due to insufficient execution.
However, the limited gains suggest that reachability alone is insufficient to detect subtle bugs when existing assertions do not adequately constrain program behavior.
% However, the overall gains remain limited, suggesting that improving reachability alone may be insufficient to detect many subtle bugs whose incorrect behavior is not adequately constrained by existing assertions.

\paragraph{Mutation-Driven Adversarial Strengthening (Stage~II).}
We further assess the contribution of the mutation-driven adversarial strengthening stage by incorporating it on top of coverage-driven augmentation.
This stage explicitly targets semantic blind spots by synthesizing plausible but incorrect mutant patches and generating tests designed to distinguish them from correct patches.
With the second stage, SWE-ABS achieves substantial additional gains.
In particular, the number of strengthened instances increases to 251, representing 44 more instances than Stage~I alone.
Meanwhile, the average drop in resolve rate rises significantly to 14.56 percentage points, corresponding to an absolute improvement of 3.21 points over Stage~I.
These results demonstrate that mutation-driven adversarial testing effectively complements coverage enhancement by uncovering semantic weaknesses that persist even when patch-relevant code is fully exercised.

\paragraph{Complementarity Analysis.}
Taken together, the ablation results suggest that coverage- and mutation-driven augmentations operate along \emph{complementary dimensions} of test adequacy.
Coverage augmentation primarily enforces execution over structurally relevant code regions, improving structural adequacy, while mutation-driven augmentation exposes semantically incorrect behaviors that remain executable yet violate intended program semantics.

\section{Conclusions, Limitations, and Future Work}

% We present a general framework for adversarial benchmark strengthening that addresses a fundamental challenge in code-generation evaluation: ensuring that test suites have sufficient discriminative power to distinguish correct from incorrect solutions.
% By systematically augmenting test suites through coverage-driven test generation and mutation-driven adversarial strengthening, our approach provides a principled methodology for improving evaluation reliability across any domain with executable test suites.
% Demonstrated on program repair benchmarks, our framework exposes that 19.78\% of patches accepted by original SWE-Bench Verified tests are semantically incorrect, strengthens 50.2\% of instances (a $25.1\times$ improvement over prior work), and causes the top-ranked agent to drop from 78.80 percentage points to 62.20 percentage points.
% Cross-benchmark evaluation further reveals that task difficulty and test strength are orthogonal: harder benchmarks do not automatically ensure stronger evaluation.
% These findings suggest that current leaderboard rankings may not accurately reflect true system capabilities, and that principled test strengthening is essential for reliable progress measurement in AI code generation.

% We finnaly choose simplified conclusion
We present SWE-ABS, a framework for adversarial benchmark strengthening that systematically augments test suites through coverage-driven generation and mutation-driven strengthening.
Applied to SWE-Bench Verified, our approach reveals that 19.78\% of accepted patches are semantically incorrect, strengthens 50.2\% of instances ($25.1\times$ over prior work), and reduces the top agent's score from 78.80\% to 62.20\%.
Cross-benchmark analysis further demonstrates that task difficulty and test strength are orthogonal, suggesting current leaderboard rankings may not reflect true code agent capabilities.

Our approach has several limitations:
(1)~\textit{Test overfitting risk}: augmented tests may encode gold-patch-specific behaviors, causing a 10.6\% false-negative rate;
(2)~\textit{Analysis scope}: intraprocedural slicing may miss cross-module dependencies;
(3)~\textit{Gold patch dependency}: requiring reference implementations limits applicability to scenarios without them.
We discuss these limitations in detail in Appendix~\ref{app:threats}.

Looking ahead, we identify several promising directions:
(1)~extending language coverage to Java, C++, and Rust through language-specific AST parsers;
(2)~integrating SWE-ABS into continuous benchmark co-evolution pipelines that periodically re-strengthen test suites;
and (3)~leveraging strengthened test suites as training signals in environments such as SWE-Gym~\cite{swe_gym}.

We believe SWE-ABS establishes a principled methodology for adversarial benchmark strengthening in code generation and repair evaluation, with potential applicability to other domains with executable oracles.

\section*{Software and Data}

All data, enhanced test suites, and evaluation scripts are publicly available at:
\url{https://github.com/OpenAgentEval/SWE-ABS}.

\section*{Impact Statement}

This work addresses a fundamental challenge in evaluating AI code generation systems: ensuring that benchmarks provide reliable signals of system capabilities rather than inflated success rates that obscure true performance.

\textbf{Reliable Evaluation Enables Informed Decisions.}
By demonstrating that 19.78\% of patches previously labeled as ``solved'' are semantically incorrect, we reveal that weak test suites can mislead both researchers prioritizing future directions and practitioners making deployment decisions. Our cross-benchmark analysis further shows that task difficulty and test strength are orthogonal: harder benchmarks do not automatically provide more discriminative evaluations.
This underscores the need for principled test strengthening as a distinct concern.

\textbf{Implications for AI Safety.}
As AI coding assistants are increasingly deployed in production environments, rigorous evaluation becomes a safety concern. Patches that pass weak tests may omit critical input validation or error handling, potentially introducing vulnerabilities when deployed. Strengthened benchmarks help surface such failure modes before real-world deployment.

\textbf{Broader Applicability.}
While demonstrated on program repair, the adversarial strengthening methodology applies to any domain where correctness can be verified through automated testing, including code translation, text-to-SQL, and robotic control. As AI systems are increasingly evaluated via test-based benchmarks, ensuring sufficient discriminative power becomes essential for meaningful progress measurement.

% \input{07Limitations}
% \input{08Ethics}
% \input{09ACK}

% -----------------------------
% Bibliography
% -----------------------------
\bibliography{references}

@inproceedings{swebench,
  title     = {{SWE}-bench: Can Language Models Resolve Real-world Github Issues?},
  author    = {Jimenez, Carlos E. and Yang, John and Wettig, Alexander and Yao, Shunyu and Pei, Kexin and Press, Ofir and Narasimhan, Karthik R.},
  booktitle = {ICLR},
  year      = {2024}
}

@inproceedings{swtbench,
  title     = {{SWT}-Bench: Testing and Validating Real-World Bug-Fixes with Code Agents},
  author    = {M{\"u}ndler, Niels and M{\"u}ller, Mark Niklas and He, Jingxuan and Vechev, Martin},
  booktitle = {NeurIPS},
  year      = {2024}
}

@article{swebench_pro,
      title={{SWE}-Bench Pro: Can AI Agents Solve Long-Horizon Software Engineering Tasks?},
      author={Deng, Xiang and Da, Jeff and Pan, Edwin and He, Yannis Yiming and Ide, Charles and Garg, Kanak and Lauffer, Niklas and Park, Andrew and Pasari, Nitin and Rane, Chetan and Sampath, Karmini and Krishnan, Maya and Kundurthy, Srivatsa and Hendryx, Sean and Wang, Zifan and Bharadwaj, Vijay and Holm, Jeff and Aluri, Raja and Zhang, Chen Bo Calvin and Jacobson, Noah and Liu, Bing and Kenstler, Brad},
  journal={arXiv preprint arXiv:2509.16941},
  year={2025}
}

@inproceedings{
multi_swebench,
title={Multi-{SWE}-bench: A Multilingual Benchmark for Issue Resolving},
author={Zan, Daoguang and Huang, Zhirong and Liu, Wei and Chen, Hanwu and Xin, Shulin and Zhang, Linhao and Liu, Qi and Li, Aoyan and Chen, Lu and Zhong, Xiaojian and Liu, Siyao and Xiao, Yongsheng and Chen, Liangqiang and Zhang, Yuyu and Su, Jing and Liu, Tianyu and Long, Rui and Ding, Ming and Xiang, Liang},
booktitle={The Thirty-ninth Annual Conference on Neural Information Processing Systems Datasets and Benchmarks Track},
year={2025}
}

@article{swebench_live,
  title={{SWE}-bench Goes Live!},
  author={Zhang, Linghao and He, Shilin and Zhang, Chaoyun and Kang, Yu and Li, Bowen and Xie, Chengxing and Wang, Junhao and Wang, Maoquan and Huang, Yufan and Fu, Shengyu and Nallipogu, Elsie and Lin, Qingwei and Dang, Yingnong and Rajmohan, Saravan and Zhang, Dongmei},
  journal={arXiv preprint arXiv:2505.23419},
  year={2025}
}

@misc{swebench_verified,
  title={Introducing {SWE}-bench Verified},
  author={Chowdhury, Neil and Aung, James and Shern, Chan Jun and Jaffe, Oliver and Sherburn, Dane and Starace, Giulio and Mays, Evan and Dias, Rachel and Aljubeh, Marwan and Glaese, Mia and Jimenez, Carlos E. and Yang, John and Ho, Leyton and Patwardhan, Tejal and Liu, Kevin and Madry, Aleksander},
  year={2024},
  url={https://openai.com/index/introducing-swe-bench-verified/},
}

@article{humaneval,
      title={Evaluating Large Language Models Trained on Code},
      author={Chen, Mark and Tworek, Jerry and Jun, Heewoo and Yuan, Qiming and Ponde de Oliveira Pinto, Henrique and Kaplan, Jared and Edwards, Harri and Burda, Yuri and Joseph, Nicholas and Brockman, Greg and Ray, Alex and Puri, Raul and Krueger, Gretchen and Petrov, Michael and Khlaaf, Heidy and Sastry, Girish and Mishkin, Pamela and Chan, Brooke and Gray, Scott and Ryder, Nick and Pavlov, Mikhail and Power, Alethea and Kaiser, Lukasz and Bavarian, Mohammad and Winter, Clemens and Tillet, Philippe and Such, Felipe Petroski and Cummings, Dave and Plappert, Matthias and Chantzis, Fotios and Barnes, Elizabeth and Herbert-Voss, Ariel and Guss, William Hebgen and Nichol, Alex and Paino, Alex and Tezak, Nikolas and Tang, Jie and Babuschkin, Igor and Balaji, Suchir and Jain, Shantanu and Saunders, William and Hesse, Christopher and Carr, Andrew N. and Leike, Jan and Achiam, Josh and Misra, Vedant and Morikawa, Evan and Radford, Alec and Knight, Matthew and Brundage, Miles and Murati, Mira and Mayer, Katie and Welinder, Peter and McGrew, Bob and Amodei, Dario and McCandlish, Sam and Sutskever, Ilya and Zaremba, Wojciech},
  journal={arXiv preprint arXiv:2107.03374},
  year={2021}
}

@article{mbpp_program_synthesis,
  title   = {Program Synthesis with Large Language Models},
  author  = {Austin, Jacob and Odena, Augustus and Nye, Maxwell and Bosma, Maarten and Michalewski, Henryk and Dohan, David and Jiang, Ellen and Cai, Carrie and Terry, Michael and Le, Quoc and Sutton, Charles},
  journal = {arXiv preprint arXiv:2108.07732},
  year    = {2021}
}

@article{eval_plus,
  title={Is Your Code Generated by Chat{GPT} Really Correct? Rigorous Evaluation of Large Language Models for Code Generation},
  author={Liu, Jiawei and Xia, Chunqiu Steven and Wang, Yuyao and Zhang, Lingming},
  journal={NeurIPS},
  year={2023}
}

@inproceedings{swe_gym,
title={Training Software Engineering Agents and Verifiers with {SWE}-Gym},
author={Pan, Jiayi and Wang, Xingyao and Neubig, Graham and Jaitly, Navdeep and Ji, Heng and Suhr, Alane and Zhang, Yizhe},
booktitle={ICML},
year={2025}
}

@inproceedings{utboost,
    title = "{UTB}oost: Rigorous Evaluation of Coding Agents on {SWE}-Bench",
    author = "Yu, Boxi  and
      Zhu, Yuxuan  and
      He, Pinjia  and
      Kang, Daniel",
    booktitle = "Proceedings of the 63rd Annual Meeting of the Association for Computational Linguistics (Volume 1: Long Papers)",
    year = "2025",
}

@article{mutation_testing,
  title={Hints on Test Data Selection: Help for the Practicing Programmer},
  author={DeMillo, Richard A. and Lipton, Richard J. and Sayward, Frederick G.},
  journal={Computer},
  volume={11},
  number={4},
  pages={34--41},
  year={1978},
  publisher={IEEE}
}

@article{mutation_testing_survey,
  title={An Analysis and Survey of the Development of Mutation Testing},
  author={Jia, Yue and Harman, Mark},
  journal={IEEE Transactions on Software Engineering},
  volume={37},
  number={5},
  pages={649--678},
  year={2011},
  publisher={IEEE}
}

@inproceedings{genprog_overfitting,
  title     = {Is the Cure Worse Than the Disease? Overﬁtting in Automated Program Repair},
  author    = {Smith, Edward K. and Barr, Earl T. and Le Goues, Claire and Brun, Yuriy},
  booktitle = {Proceedings of the 2015 10th Joint Meeting on Foundations of Software Engineering},
  year      = {2015}
}

@inproceedings{patch_plausibility_and_correctness,
  title     = {An Analysis of Patch Plausibility and Correctness for Generate-and-Validate Patch Generation Systems},
  author    = {Qi, Zichao and Long, Fan and Achour, Sara and Rinard, Martin},
  booktitle = {Proceedings of the 2015 International Symposium on Software Testing and Analysis},
  year      = {2015},
}

@article{mutation_testing_study_llm,
  title         = {A Comprehensive Study on Large Language Models for Mutation Testing},
  author        = {Wang, Bo and Chen, Mingda and Deng, Ming and Lin, Youfang and Harman, Mark and Papadakis, Mike and Zhang, Jie M.},
  journal = {arXiv preprint arXiv:2406.09843},
  year    = {2026}
}

@inproceedings{mutation_guided_llm_based_test_generation_at_meta,
  title     = {Mutation-Guided LLM-based Test Generation at Meta},
  author    = {Harman, Mark and Ritchey, Jillian and Harper, Inna and Sengupta, Shubho and Mao, Ke and Gulati, Abhishek and Foster, Christopher and Robert, Herv\'{e}},
  booktitle = {Proceedings of the 33rd ACM International Conference on the Foundations of Software Engineering},
  year      = {2025},
}

@misc{coveragepy,
  author = {Batchelder, Ned},
  title  = {{Coverage.py: The code coverage tool for Python}},
  url    = {https://github.com/coveragepy/coveragepy},
  year   = {2026},
  note   = {version 7.13.2}
}

@misc{go-coverage,
    author = {Pike, Rob},
    title = {The cover story},
    year = {2013},
    url = {https://go.dev/blog/cover}
}

@misc{istanbuljs-nyc,
  author = {Istanbuljs},
  title = {nyc: the Istanbul command line interface},
  year = {2024},
  url = {https://github.com/istanbuljs/nyc},
}

@misc{v8-coverage,
  author = {{V8 Developers}},
  title = {JavaScript code coverage},
  year = {2017},
  howpublished = {\url{https://v8.dev/blog/javascript-code-coverage}},
}

@misc{tree-sitter,
    author = {Tree-sitter},
    title = {Tree-sitter is a parser generator tool and an incremental parsing library},
    year = {2026}
}

@ARTICLE{2020SciPy-NMeth,
  author  = {Virtanen, Pauli and Gommers, Ralf and Oliphant, Travis E. and
            Haberland, Matt and Reddy, Tyler and Cournapeau, David and
            Burovski, Evgeni and Peterson, Pearu and Weckesser, Warren and
            Bright, Jonathan and {van der Walt}, St{\'e}fan J. and
            Brett, Matthew and Wilson, Joshua and Millman, K. Jarrod and
            Mayorov, Nikolay and Nelson, Andrew R. J. and Jones, Eric and
            Kern, Robert and Larson, Eric and Carey, C J and
            Polat, {\.I}lhan and Feng, Yu and Moore, Eric W. and
            {VanderPlas}, Jake and Laxalde, Denis and Perktold, Josef and
            Cimrman, Robert and Henriksen, Ian and Quintero, E. A. and
            Harris, Charles R. and Archibald, Anne M. and
            Ribeiro, Ant{\^o}nio H. and Pedregosa, Fabian and
            {van Mulbregt}, Paul and {SciPy 1.0 Contributors}},
  title   = {{{SciPy} 1.0: Fundamental Algorithms for Scientific
            Computing in Python}},
  journal = {Nature Methods},
  year    = {2020},
  volume  = {17},
  pages   = {261--272},
  adsurl  = {https://rdcu.be/b08Wh},
  doi     = {10.1038/s41592-019-0686-2},
}

@article{trae_agent,
      title={Trae Agent: An LLM-based Agent for Software Engineering with Test-time Scaling}, 
      author={{Trae Research Team} and Gao, Pengfei and Tian, Zhao and Meng, Xiangxin and Wang, Xinchen and Hu, Ruida and Xiao, Yuanan and Liu, Yizhou and Zhang, Zhao and Chen, Junjie and Gao, Cuiyun and Lin, Yun and Xiong, Yingfei and Peng, Chao and Liu, Xia},
  journal={arXiv preprint arXiv:2507.23370},
  year={2025}
}

@article{seed_coder,
    title={{Seed-Coder}: Let the Code Model Curate Data for Itself},
    author={{ByteDance Seed} and Zhang, Yuyu and Su, Jing and Sun, Yifan and Xi, Chenguang and Xiao, Xia and Zheng, Shen and Zhang, Anxiang and Liu, Kaibo and Zan, Daoguang and Sun, Tao and Zhu, Jinhua and Xin, Shulin and Huang, Dong and Bai, Yetao and Dong, Lixin and Li, Chao and Chen, Jianchong and Zhou, Hanzhi and Huang, Yifan and Ning, Guanghan and Song, Xierui and Chen, Jiaze and Liu, Siyao and Shen, Kai and Xiang, Liang and Wu, Yonghui}, 
    journal={arXiv preprint arXiv:2506.03524},
    year={2025}
}

@article{live_swe_agent,
  title={Live-SWE-agent: Can Software Engineering Agents Self-Evolve on the Fly?},
  author={Xia, Chunqiu Steven and Wang, Zhe and Yang, Yan and Wei, Yuxiang and Zhang, Lingming},
  journal={arXiv preprint arXiv:2511.13646},
  year={2025}
}

@article{gemini,
      title={Gemini: A Family of Highly Capable Multimodal Models}, 
      author={{Gemini Team}},
    journal={arXiv preprint arXiv:2312.11805},
    year={2025}
}

@misc{acoder,
  author       = {{ACoder Team} and Yang, Xikai and Teng, Tengjixiang and Zhang, Xinhe and He, Jiawei and Zhou, Changkong and Qiao, Zhidong and Zhang, Xi and Zou, Tao},
  title        = {ACoder - Achieving State-of-the-Art Performance on SWE-bench Verified},
  year         = {2025},
  url ={https://github.com/ACoder-AI/ACoder/blob/main/ACoder_Report.md},
}

@misc{atlassian_rovo_dev,
    title = {Rovo Dev | Agentic AI for software teams | Atlassian},
    url = {https://www.atlassian.com/software/rovo-dev},
    author = {Atlassian},
    year = {2025}
}

@misc{epam_ai_run_developer_agent,
  title                    = {Strategic AI Development Services for Enterprises | EPAM},
  url                      = {https://www.epam.com/services/artificial-intelligence},
  author                   = {{EPAM Systems, Inc.}},
  year                     = {2025}
}

@misc{glm-47,
      title={GLM-4.5: Agentic, Reasoning, and Coding (ARC) Foundation Models}, 
      author={{GLM Team}},
      year={2025},
      eprint={2508.06471},
      archivePrefix={arXiv},
      primaryClass={cs.CL},
      url={https://arxiv.org/abs/2508.06471}, 
}

@misc{gpt-5,
      title={OpenAI GPT-5 System Card}, 
      author={OpenAI},
      year={2025},
      eprint={2601.03267},
      archivePrefix={arXiv},
      primaryClass={cs.CL}
}

@article{frontiercs,
  title={FrontierCS: Evolving Challenges for Evolving Intelligence},
  author={Mang, Qiuyang and Chai, Wenhao and Li, Zhifei and Mao, Huanzhi and Zhou, Shang and Du, Alexander and Li, Hanchen and Liu, Shu and Chen, Edwin and Wang, Yichuan and others},
  journal={arXiv preprint arXiv:2512.15699},
  year={2025}
}
\bibliographystyle{icml2026}

% -----------------------------
% Appendix
% -----------------------------
\newpage
\appendix
\onecolumn

% Reset table and figure counters for appendix and use section prefix
% \setcounter{table}{0}
% \setcounter{figure}{0}
% \renewcommand{\thetable}{A\arabic{table}}
% \renewcommand{\thefigure}{A\arabic{figure}}

% \input{A0_Appendix}
\clearpage
\onecolumn
\section{Appendix}

The appendix is organized as follows:
\begin{itemize}
    \item \textbf{Appendix~\ref{app:prompts}} lists all prompt templates used in the pipeline.
    \item \textbf{Appendix~\ref{app:test_decoupling_cases}} provides test decoupling case studies.
    \item \textbf{Appendix~\ref{app:setup}} details experimental implementation including hyperparameters and program slicing.
    \item \textbf{Appendix~\ref{app:mutation_case}} covers mutation cases and evaluation.
    \item \textbf{Appendix~\ref{app:full_performance}} provides full leaderboard results for top-30 agents on SWE-Bench Verified.
    \item \textbf{Appendix~\ref{app:Bash_Only_full_performance}} provides full leaderboard results for Bash-only agents on SWE-Bench Verified.
    \item \textbf{Appendix~\ref{app:results}} reports additional experimental results by repository and language.
    \item \textbf{Appendix~\ref{app:cost}} presents computational cost analysis.
    \item \textbf{Appendix~\ref{app:generalization}} presents generalization analysis.
    \item \textbf{Appendix~\ref{app:test_quality}} presents test quality analysis.
    \item \textbf{Appendix~\ref{app:cases}} presents qualitative case studies of test strengthening failures and agent-generated patch failures.
    \item \textbf{Appendix~\ref{app:threats}} discusses limitations and threats to validity.
\end{itemize}

% ===== Prompts (Method 3.1.1) =====

\subsection{Prompt Templates}
\label{app:prompts}

\subsubsection{stage~I: Initial Test Generation Prompt}

\begin{lstlisting}[style=promptstyle]
You are a helpful assistant that can interact multiple times with a computer shell to solve programming tasks.
Your response must contain exactly ONE bash code block with ONE command (or commands connected with && or ||).

Include a THOUGHT section before your command where you explain your reasoning process.
Format your response as shown in <format_example>.

<format_example>
THOUGHT: Your reasoning and analysis here

```bash
your_command_here
```
</format_example>

Failure to follow these rules will cause your response to be rejected.

<pr_description>
Consider the following PR description:
{{task}}
</pr_description>

<instructions>
# Task Instructions

## Overview
You're a software engineer interacting continuously with a computer by submitting commands.
Your task is to reproduces the issue by writing independent test case and cover as many corner cases as possible.

There is a gold patch that is used to fix the issue.
--- Gold Patch ---
```
{{gold_patch}}
```

There is an original test patch that is used to verify the fix. You can learn from it to generate new test cases that thoroughly test the fix.
--- Original Test Patch ---
```
{{test_patch}}
```

IMPORTANT: 
- This is an interactive process where you will think and issue ONE command, see its result, then think and issue your next command.
- {{workdir}} is the working directory for all your subsequent commands.
- Gold patch is already apply to the given repository.
- Test execution command example in this repository: {{test_command}}
- Write the new test case in a new test file, do not modify the original test file!!!
- Do not use other test command, do not run the full test suite

## Recommended Workflow
1. Understand the Problem:  
  (1) Begin by carefully reading the user's problem description to fully grasp the issue. 
  (2) Identify the core components and expected behavior. 
2. Explore and Locate:  
  (1) Use the available tools to explore the repository structure and locate the source and test directories.
  (2) Based on the gold patch and the original test patch, identify the most relevant source files and test files, and analyze how the original test verifies the fix.
3. Expand Test Coverage:  
  (1) Extend the original test into a broader test suite, covering edge cases, corner cases, and stress conditions. 
  (2) When writing tests, ensure generality and avoid overfitting assertions to the specific implementation of the gold patch.
      - Focus on verifying the correctness of behavior, not on matching the exact output form of the gold patch.
  (3) Run the tests to confirm that all pass with the gold patch applied.
  (4) If any tests fail, adjust the inputs or expected outputs until a stable and correct test suite is achieved.  
4. Finalize Test Suite:
  (1) Consolidate both the original and extended tests into a single new test file, rather than providing a git diff.
  
\end{lstlisting}

\subsubsection{stage~I: Test Decoupling Prompt}

\begin{lstlisting}[style=promptstyle]
Now you must review the tests you have written:

1. Check whether there are any hard-coded assertions on error messages. 
  If so, determine whether the assertion relies on keywords, fields, or formatting specifics introduced by the gold patch; 
  if it does, replace it with a more general assertion that verifies only the essential error semantics rather than exact strings.

2. Check whether any test enforces strict ordering when the Issue does not require ordering.  
  If so, revise the test to avoid relying on order.

3. Check for any other situations where the test is too closely tied to the gold patch implementation
  If so, adjust to avoid unnecessary coupling.

Do not create new test files again. If you believe any of the new tests you previously wrote may suffer from the issues above, revise them directly.
Do not install any additional libraries. If your test files import libraries that are not available in the current environment, replace them with alternatives that are available.

## Testing
After making the necessary modifications, run your tests again using:
{{test_command}}

IMPORTANT! Execute the test_command by itself. Do not combine it with any other commands.

If any tests fail, adjust the inputs or expected outputs until a stable and correct test suite is achieved.

## Submission
Before submitting, you MUST execute the following Git command to inspect the modified files.
You MUST use this command (or an equivalent one that only lists file names and does not show diffs):

```bash
git status --short
```
Based on the output of this command, you MUST remove any unnecessary or temporary generated files
(e.g., cache files, logs, build artifacts, or intermediate outputs).

You MUST repeat the file-listing step if files are removed, until only intended changes remain.

Only after completing the cleanup and when no further progress can be made,
you may issue EXACTLY the following command.

IMPORTANT:
The final submission command MUST be sent as a standalone instruction and MUST NOT be combined with any other instructions.

```bash
cd {{workdir}} && echo COMPLETE_TASK_AND_SUBMIT_FINAL_OUTPUT && git add -A && git diff --cached
```

# Workflow (You must follow this workflow):
  1. Review existing tests for hard-coded assertions, ordering assumptions, and tight coupling to the gold patch.
  2. (Optional) Revise the problematic tests directly to make them more general and robust.
  3. Run the test suite using the provided test command and fix any remaining failures.
  4. Once all tests pass and no further improvements are possible, submit the final changes.
\end{lstlisting}

\subsubsection{stage~I: Coverage-Guided Augmentation Prompt}

\begin{lstlisting}[style=promptstyle]
The test you wrote covers {{coverage_rate}} of the lines modified by the gold patch.
Now you need to update your test cases to cover these lines of code.
Do not create new test files again.
Miss line:
{{error_info}}

## Testing
After making the necessary modifications, run your tests again using:
{{test_command}}

If any tests fail, adjust the inputs or expected outputs until a stable and correct test suite is achieved.

## Submission
Before submitting, you MUST execute the following Git command to inspect the modified files.
You MUST use this command (or an equivalent one that only lists file names and does not show diffs):

```bash
git status --short
```
Based on the output of this command, you MUST remove any unnecessary or temporary generated files
(e.g., cache files, logs, build artifacts, or intermediate outputs).

You MUST repeat the file-listing step if files are removed, until only intended changes remain.

Only after completing the cleanup and when no further progress can be made,
you may issue EXACTLY the following command.

IMPORTANT:
The final submission command MUST be sent as a standalone instruction and MUST NOT be combined with any other instructions.

```bash
cd {{workdir}} && echo COMPLETE_TASK_AND_SUBMIT_FINAL_OUTPUT && git add -A && git diff --cached
```

# Workflow (You must follow this workflow):
  1. Identify the missed lines reported in the coverage feedback.
  2. Update existing test cases to exercise the uncovered code paths.
  3. Run the test command and verify that coverage improves and all tests pass.
  4. Iterate on test inputs or expectations until the test suite is stable.
  5. Submit the final changes.
\end{lstlisting}

\subsubsection{stage~II: Mutant Generation Prompt}

\begin{lstlisting}[style=promptstyle]
You are a helpful assistant that can interact multiple times with a computer shell to solve programming tasks.
Your response must contain exactly ONE bash code block with ONE command (or commands connected with && or ||).

Include a THOUGHT section before your command where you explain your reasoning process.
Format your response as shown in <format_example>.

<format_example>
THOUGHT: Your reasoning and analysis here

```bash
your_command_here
```
</format_example>

Failure to follow these rules will cause your response to be rejected.

<pr_description>
Consider the following PR description:
{{task}}
</pr_description>

<instructions>
# Task Instructions

## Overview
You are an autonomous software engineer interacting with a live repository.
Your job is to synthesize a mutation patch that:

1. Alters program behavior in subtle but semantically meaningful ways  
  (NOT formatting or refactoring - but true behavior deviations)

2. Still passes ALL tests defined in the given Test Patch

3. Does NOT implement the correct fix described in the Gold Patch  
  (Your mutation must be wrong but test-passing)

This creates a *surviving mutant* that exposes weaknesses in the existing tests.


There is a gold patch that is used to fix the issue.
--- Gold Patch ---
```
{{gold_patch}}
```

There is an test patch that is used to verify the fix.
---Test Patch ---
```
{{test_patch}}
```
Test execution command: {{test_command}}

IMPORTANT: 
- This is an interactive process where you will think and issue ONE command, see its result, then think and issue your next command.
- {{workdir}}  is the working directory for all your subsequent commands.
- Test Patch and Gold Patch is already apply to the given repository.
- Do not run the full test suite, just run the test in the test patch
- Do NOT run git commit

## Hard Requirements for Your Mutation
1. Directly modify the repository source code
2. Mutation must change semantics (examples):
  - Modify branching logic
  - Alter returned values
  - Skip a step of processing
  - Incorrect edge case handling
  - Loosen a condition
3.Your mutation MUST be logically connected to the original issue
3. Your mutation MUST cause behavior divergence relative to the gold patch  
  (i.e., gold_patch -> correct; your_patch -> subtly incorrect behavior)
4. After mutation, the repository MUST PASS all tests in the Test Patch
5. Absolutely forbidden:
  - Comment-only or formatting changes
  - Pure refactoring with identical semantics
  - Dead code insertion
  - Re-implementing the gold patch
6. Before writing your mutation, explicitly state:
  **Why this mutation is semantically wrong but can evade the tests.**

## Required Workflow
1. Establish baseline test status 
  - Execute the test command on the current codebase  
  - Record all failing test cases as baseline failures
  - These failures are considered acceptable and should NOT be fixed  
  - Any additional failures introduced later must be eliminated  

2. Understand the issue
  - Read PR description + gold patch + test patch  
  - Identify the core expected behavior  
  - Infer what aspects the test suite fails to validate or under-specifies  

3. Locate files 
  - Use commands to search the codebase  
  - Identify where the gold patch would apply  
  - Locate alternative or related code paths implementing similar logic  

4. Design an adversarial mutation  
  - Intentionally introduce a logically incorrect or weakened behavior  
  - The mutation may preserve baseline failures  
  - The mutation must NOT introduce new failing test cases  

5. Iteratively refine
  - Apply the mutation and execute the tests  
  - Compare failing tests against the baseline failures  
  - If new test failures are observed -> adjust the mutation  
  - Continue until a surviving mutant is produced with no additional failures  

6. Submit the final mutation
  - Include a justification of:  
    - What semantic behavior was altered or broken  
    - Why the existing tests cannot catch this change  

\end{lstlisting}

\subsubsection{stage~II: Relevance and Equivalence Filtering Prompt}

\begin{lstlisting}[style=promptstyle]
You are a highly experienced software developer tasked with evaluating the quality of a Mutation in a codebase.

You will receive the following four parts:

1. GitHub Issue Description - Background of the problem and the fix objective;
2. Gold Patch - The official or correct fix patch;
3. Test Patch - The tests added to verify the fix;
4. Mutation - A small code change introduced based on the gold patch, which has passed all the new tests.

Your task is to evaluate whether this Mutation is a high-quality surviving mutant.

#### Step 1: Difference Identification (Difference Analysis)

First, compare the Mutation and Gold Patch,
Summarize their differences (such as added/removed/modified logic, comments, condition checks, function calls, etc.).
The output should include:

 - Modified Lines (you can reference the key context)
 - Change Type (Added/Modified/Deleted)
 - A brief explanation of the semantic change or lack thereof (e.g., "only a comment difference," "condition branch logic adjustment")

#### Step 2: In-depth Evaluation (Structured Evaluation)

1. Relevance to the Issue (Relevance)
  - Does the Mutation modify the part of the code related to the Issue fix?
  - Does it still fall within the scope of the Issue's logic (e.g., the same function, path, or input conditions)?
  - If it is a mutation to a parameter, does the issue explicitly or implicitly specify the original value of that parameter? If not, then it is irrelevant.
  - If it's unrelated (e.g., just style or comment changes), its relevance is low.


2. Does it constitute a valid mutation? (Mutation Validity)
  - Does the **Mutation** introduce a semantic difference compared to the **Gold Patch**?
  - Is this change likely to alter behavior in certain scenarios and potentially cause a functional bug, or is it an equivalent mutation?
  - Can this difference reveal a blind spot in the test suite?
  - A high-quality surviving mutant should meet the following criteria:
    * The functionality is mostly correct, but with slight semantic deviation.
    * The test suite did not detect it.
    * It provides insight into potential test improvements.


### Output Format (Please Follow This Exactly)
Difference Summary:
  Briefly list all the logical differences between the **Gold Patch** and the **Mutation**.

Reasoning Analysis:
  Relevance to the Issue:
  Mutation Validity (Does it truly constitute a mutation?):
  

Final Evaluation: Please output the final judgment in the following format, placed inside the <Answer></Answer> tag:
<Answer>
Relevance: Yes/No
Mutation Validity: Yes/No
</Answer>

### Input Format

```
## Issue
{{issue}}

## Test Patch
{{test_patch}}

## Gold Patch
{{gold_patch}}

## Mutation
{{mutation}}
```

\end{lstlisting}

\subsubsection{Stage II: Prompt for Mutant-Guided Test Augmentation of Non-Equivalent Mutants}

\begin{lstlisting}[style=promptstyle]
You are a helpful assistant that can interact multiple times with a computer shell to solve programming tasks.
Your response must contain exactly ONE bash code block with ONE command (or commands connected with && or ||).

Include a THOUGHT section before your command where you explain your reasoning process.
Format your response as shown in <format_example>.

<format_example>
THOUGHT: Your reasoning and analysis here

```bash
your_command_here
```
</format_example>

Failure to follow these rules will cause your response to be rejected.

<pr_description>
Consider the following PR description:
{{task}}
</pr_description>

<instructions>
# Task Instructions

## Overview
You're a software engineer interacting continuously with a computer by submitting commands.
Your task is to directly modify and enhance the current test code to ensure that logic deviations can be detected.  

You are given three code snippets related to a specific GitHub issue:
There is a gold patch that is used to fix the issue.
1. Gold Patch - the correct fix for the issue.  
  ---  
  {{gold_patch}}  
  ---  

2. Mutation Patch - a mutated version of the gold patch that still passes the existing tests (a surviving mutant).  
  ---  
  {{mutation_patch}}  
  ---  

3. Mutation thinking derived from mutations in the Gold Patch.
  ---
  {{mutation_thinking}}
  ---

4. Test Patch - the existing test code that was originally used to verify the gold patch.  
  ---  
  {{test_patch}}  
  --- 
Test execution in this repository: {{test_command}}


Your goal is to **enhance the given test code** (from the Test Patch) so that:  
- The **Gold Patch** still passes all tests (it represents the correct behavior).  
- The **Mutation Patch** fails at least one test (it represents an incorrect but surviving variant). 

Requirements for the new test:
- Modify only the test code.  
- Add minimal but **targeted** assertions or test cases that can expose the behavioral difference between the Gold Patch and the Mutation Patch.  

IMPORTANT: 
- This is an interactive process where you will think and issue ONE command, see its result, then think and issue your next command.
- {{workdir}}  is the working directory for all your subsequent commands.
- You will be working with two repositories: **Gold** and **Mutated**, which have respectively applied the Gold Patch and the Mutation Patch.  
- The Test Patch has already been applied to both repositories.  
- Do not run the full test suite, just run the test in the test patch
- Do not change the test command, only use the given test command
- When writing commands, always specify which repository you are operating on. For example: `<env>Gold</env> cd testbed && echo 1` or `<env>Mutated</env> cd testbed && echo 1`.
- If you want to run the command in both repositories, you can use the `<env>All</env>` tag. For example: `<env>All</env> cd testbed && echo 1` will run `cd testbed && echo 1` in both Gold and Mutated repositories.
- If you use the <env>All</env>tag, do not include any other environment-specific tags afterward, as it will cause a parsing error.


## Recommended Workflow
1. Understand the Problem:  
  (1) Begin by carefully reading the user's problem description to fully grasp the issue. 
  (2) Identify the core components and expected behavior. 
2. Explore and Locate:  
  (1) Use the available tools to explore the repository structure and locate the source and test directories.
  (2) Based on the gold patch and the test patch, identify the most relevant source files and test files
  (3) Identify the diffence between the gold patch and the mutation patch
3. Modify the test file and debug:   
  (1) Modify and enhance the current test code in test_patch
  (2) Debug the test code to ensure it passes the gold patch and fails the mutation patch
  (3) If any tests fail with gold patch, adjust the inputs or expected outputs until a stable and correct test suite is achieved.
  (4) Ensure that the newly added test cases are also general, rather than being gold-patch-specific.
4. Submit
  (1) Before you submit, run the given tests command to ensure that all tests pass with the gold patch applied and fail with the mutation patch applied.


\end{lstlisting}

\subsubsection{Stage II: Prompt for Mutant-Guided Test Augmentation of Equivalent Mutants}

\begin{lstlisting}[style=promptstyle]

<pr_description>
Consider the following PR description:
{{task}}
</pr_description>

<instructions>
# Task Instructions

## Overview
You're a software engineer interacting continuously with a computer by submitting commands.
Your task is to directly modify the current test code to ensure that the equivalent mutation (equ_mutation) can pass.

You are given three code snippets related to a specific GitHub issue:
1. Gold Patch - the correct fix for the issue.  
  ---  
  {{gold_patch}}  
  ---  

2. Mutation Patch - a equivalent mutated version of the gold patch that fails the existing tests.  
  ---  
  {{mutation_patch}}  
  ---  

3. Test Patch - the existing test code that was originally used to verify the gold patch.  
  ---  
  {{test_patch}}  
  --- 
Test execution in this repository: {{test_command}}


Your goal is to **modify the given test code** (from the Test Patch) so that:  
- The **Gold Patch** and **Mutation Patch** passes all tests (it represents the correct behavior).   

Requirements for the new test:
- Modify only the test code.  

IMPORTANT: 
- This is an interactive process where you will think and issue ONE command, see its result, then think and issue your next command.
- /testbed  is the working directory for all your subsequent commands.
- You will be working with two repositories: **Gold** and **Mutated**, which have respectively applied the Gold Patch and the Mutation Patch.  
- The Test Patch has already been applied to both repositories.  
- Do not run the full test suite, just run the test in the test patch
- Do not change the test command, only use the given test command
- When writing commands, always specify which repository you are operating on. For example: `<env>Gold</env> cd testbed && echo 1` or `<env>Mutated</env> cd testbed && echo 1`.
- If you want to run the command in both repositories, you can use the `<env>All</env>` tag. For example: `<env>All</env> cd testbed && echo 1` will run `cd testbed && echo 1` in both Gold and Mutated repositories.
- If you use the <env>All</env>tag, do not include any other environment-specific tags afterward, as it will cause a parsing error.


## Recommended Workflow
0. Check if it is an equivalent mutation:
  (1) If it is not an equivalent mutation, submit directly.
1. Understand the Problem:  
  (1) Begin by carefully reading the user's problem description to fully grasp the issue. 
  (2) Identify the core components and expected behavior. 
2. Explore and Locate:  
  (1) Use the available tools to explore the repository structure and locate the source and test directories.
  (2) Based on the gold patch and the test patch, identify the most relevant source files and test files
  (3) Identify the parts of the Test Patch that cause the mutation patch to fail.
3. Modify the test file and debug:   
  (1) Modify and enhance the current test code in test_patch
  (2) Debug the test code to ensure it passes the gold patch and the mutation patch
  (3) If any tests fail with gold patch and the mutation patch, adjust the inputs or expected outputs until a stable and correct test suite is achieved.
4. Submit
  (1) Before you submit, run the given tests command to ensure that all tests pass with the gold patch applied and mutation patch applied.

\end{lstlisting}

% ===== Test Decoupling (Method 3.1.2) =====

\subsection{Test Decoupling Case Studies}
\label{app:test_decoupling_cases}

\textbf{Instance.} astropy\_\_astropy-13033

\textbf{Issue.}
When a required column other than \texttt{time} is removed from a \texttt{TimeSeries} object, the raised exception is misleading.  
Instead of indicating that a required column is missing, the error message claims that the first column is invalid, sometimes comparing identical values (e.g., \texttt{'time'} vs. \texttt{'time'}), which confuses users.

\paragraph{Gold Fix.}
Listing~\ref{lst:gold_required_columns_fix} shows the gold patch that corrects the exception message.
Rather than hard-coding the first required column, the fix reports the full list of required columns and the actual columns found, producing a semantically accurate and less confusing error message.

\begin{lstlisting}[language=diff,caption={Gold patch: corrected error message for missing required columns},label={lst:gold_required_columns_fix}]
@@ -76,9 +83,10 @@ def _check_required_columns(self):
 
             elif self.colnames[:len(required_columns)] != required_columns:
 
-                raise ValueError("{} object is invalid - expected '{}' "
-                                 "as the first column{} but found '{}'"
-                                 .format(self.__class__.__name__, required_columns[0], plural, self.colnames[0]))
+                raise ValueError("{} object is invalid - expected {} "
+                                 "as the first column{} but found {}"
+                                 .format(self.__class__.__name__, as_scalar_or_list_str(required_columns),
+                                            plural, as_scalar_or_list_str(self.colnames[:len(required_columns)])))
\end{lstlisting}

\paragraph{Original Test Patch.}
Listing~\ref{lst:original_test_patch} presents the original test added alongside the fix.
This test validates the new behavior by asserting the exact error message string, tightly coupling the test to a specific formatting of the exception message.

\begin{lstlisting}[language=diff,caption={Original test patch asserting exact error message},label={lst:original_test_patch}]
diff --git a/astropy/timeseries/tests/test_sampled.py b/astropy/timeseries/tests/test_sampled.py
--- a/astropy/timeseries/tests/test_sampled.py
+++ b/astropy/timeseries/tests/test_sampled.py
@@ -395,6 +395,14 @@ def test_required_columns():
     assert exc.value.args[0] == ("TimeSeries object is invalid - expected "
                                  "'time' as the first column but found 'banana'")
 
+    # https://github.com/astropy/astropy/issues/13009
+    ts_2cols_required = ts.copy()
+    ts_2cols_required._required_columns = ['time', 'a']
+    with pytest.raises(ValueError) as exc:
+        ts_2cols_required.remove_column('a')
+    assert exc.value.args[0] == ("TimeSeries object is invalid - expected "
+                                 "['time', 'a'] as the first columns but found ['time', 'b']")
+
 
 @pytest.mark.parametrize('cls', [BoxLeastSquares, LombScargle])
 def test_periodogram(cls):
\end{lstlisting}

\paragraph{Initial Generated Test (Overfitted).}
Listing~\ref{lst:overfitted_test} shows an automatically generated test that further overfits to a specific error message format.
Although functionally correct, the test assumes a particular phrasing and column representation, making it fragile to legitimate refactorings of the error message.

\begin{lstlisting}[language=diff,caption={Overfitted generated test relying on exact error message},label={lst:overfitted_test}]
+def test_relax_mode_invalid_first_column_with_multiple_required():
+    # Empty initialization triggers relax mode
+    ts = TimeSeries()
+    # Set multiple required columns
+    ts._required_columns = ['time', 'a']
+    # Add wrong first column when relax=True should give single-column message
+    with pytest.raises(ValueError) as exc:
+        ts.add_column(Column([1, 2, 3], name='flux'), index=0)
+    assert exc.value.args[0] == ("TimeSeries object is invalid - expected "
+                                 "'time' as the first column but found 'flux'")
\end{lstlisting}

\paragraph{Decoupled Test.}
Listing~\ref{lst:decoupled_test} presents the improved, decoupled version of the test.
Instead of asserting the full error string, the test checks for key semantic elements (e.g., the presence of ``expected'', ``found'', and the relevant column names).
This approach preserves test robustness while still validating the intended behavior.

\begin{lstlisting}[language=diff,caption={Decoupled test asserting semantic properties of the error message},label={lst:decoupled_test}]
+def test_relax_mode_invalid_first_column_with_multiple_required():
+    ts = TimeSeries()
+    ts._required_columns = ['time', 'a']
+    with pytest.raises(ValueError) as exc:
+        ts.add_column(Column([1, 2, 3], name='flux'), index=0)
+    msg = str(exc.value)
+    assert "expected" in msg and "found" in msg
+    assert "first column" in msg
+    assert "'time'" in msg and "'flux'" in msg
\end{lstlisting}

% ===== Experimental Setup (Method 3.1.3-3.1.4) =====

\subsection{Detailed Experimental Implementation}
\label{app:setup}

\subsubsection{Hyperparameters}
\label{app:hyperparameters}

\begin{itemize}
    \item LLM temperature: 1.0 (default)
    \item Max test generation attempts: 3
    \item Max mutants generated per instance: 2
    \item Max test augmentation attempts per adversarial mutant: 3
    \item Test execution timeout: 120 seconds
    \item Max dependency graph depth: 5 hops
\end{itemize}

\subsubsection{Program Slicing Implementation}
\label{app:slicing_implementation}

We perform intraprocedural program slicing to identify code regions that are data-flow or control-flow dependent on a given patch.
We use Tree-sitter~\cite{tree-sitter} to parse source files across multiple languages, including Python, JavaScript, Go, and TypeScript.

\begin{enumerate}
    \item \textbf{AST Parsing and Scope Identification}: Parse modified source files into abstract syntax trees and build a line-to-scope mapping that identifies whether each line belongs to a function, class, or global scope. This enables scope-limited slicing to prevent excessive propagation across unrelated code regions.

    \item \textbf{Executable Line Extraction}: Identify executable statements (assignments, returns, function/class definitions, etc.) and filter out non-executable lines such as comments, docstrings, and blank lines. For multi-line statements (e.g., function signatures, method calls), we map modified lines within the statement to its starting line. We denote the set of all executable lines as $L_{\text{executable}}$.

    \item \textbf{Global Modification Filtering}: Filter out semantically insignificant global-level modifications such as import statements and simple variable assignments, retaining only modifications with substantial semantic impact (e.g., function/class definitions, control flow statements).

    \item \textbf{Def-Use Analysis}: Construct data dependency information by analyzing variable definitions (assignments) and uses (loads) at each line using an AST visitor pattern.

    \item \textbf{Bidirectional k-Hop Slicing}:
    \begin{itemize}
        \item \textbf{Forward slicing}: Starting from modified lines, propagate through def-use chains to find lines that use variables defined by the patch (def $\to$ use). We denote the forward slice as $L_{\text{fwd}}$.
        \item \textbf{Backward slicing}: Propagate through use-def chains to find lines that define variables used by the patch (use $\to$ def). We denote the backward slice as $L_{\text{bwd}}$.
        \item Use $k$-hop propagation (typically $k=1$ or $k=5$) to limit slice size while capturing relevant dependencies.
        \item Optionally restrict propagation to the same scope (function/class/global) as modified lines to avoid excessive expansion.
    \end{itemize}

    \item \textbf{Slice Union}: Combine forward and backward slices, then intersect with executable lines to obtain the final patch-relevant lines $L_{\text{rel}} = (L_{\text{fwd}} \cup L_{\text{bwd}}) \cap L_{\text{executable}}$.
\end{enumerate}

This design balances precision and scalability: scope-limited slicing prevents uncontrolled expansion, while bidirectional propagation captures both upstream dependencies (what the patch depends on) and downstream impacts (what depends on the patch).

\subsubsection{Coverage Calculation}
\label{app:coverage_calculation}

To measure the effectiveness of coverage-driven augmentation (Stage~I), we compute the percentage of patch-relevant lines $L_{\text{rel}}$ executed by a given test suite using:
\begin{equation*}
\text{Coverage}(T, L_{\text{rel}}) = \frac{|L_{\text{exec}} \cap L_{\text{rel}}|}{|L_{\text{rel}}|},
\end{equation*}
where $L_{\text{exec}}$ denotes the set of executed lines and $L_{\text{rel}}$ denotes patch-relevant lines from program slicing.
This formula applies uniformly across both benchmarks; the key difference lies in how we collect $L_{\text{exec}}$ for different programming languages and repository configurations.

\paragraph{SWE-Bench (Python).}
For SWE-Bench instances, we use \texttt{trace.py} from the SWT-Bench~\cite{swtbench} repository to collect line-level execution traces.

\paragraph{SWE-Bench Pro (Multi-Language).}
SWE-Bench Pro introduces instances in Python, JavaScript, Go, and TypeScript, requiring language-specific coverage tools.
We implement a unified coverage parsing framework that standardizes output from heterogeneous tools into a common representation:

\textbf{Python}: We use \texttt{coverage.py}~\cite{coveragepy} to collect line-level execution traces.

\textbf{Go}: We use Go's built-in coverage profiler (\texttt{go test -coverprofile})~\cite{go-coverage}, which produces line-range coverage data that we parse to extract executed lines.

\textbf{JavaScript}: We employ Istanbul/nyc~\cite{istanbuljs-nyc} to collect statement-level coverage, from which we extract executed line numbers.

\textbf{TypeScript}: We use Node.js V8 coverage~\cite{v8-coverage}, which provides byte-offset-based coverage that we convert to line numbers through exact mapping.

All language-specific parsers output a standardized \texttt{CoverageResult} structure containing executed and missing lines per file.
Once $L_{\text{exec}}$ is extracted for each language, we apply the same coverage formula uniformly across all instances.

\subsubsection{Spearman Calculation}
\label{app:spearman_calculation}

To quantify the stability of leaderboard rankings under test augmentation, we compute the Spearman rank correlation coefficient $\rho$ between the original and augmented agent rankings.

\paragraph{Mathematical Definition.}
Given $n$ agents with original ranks $r_i$ and augmented ranks $r'_i$ (where $i = 1, \ldots, n$), Spearman's $\rho$ is defined as:
\begin{equation*}
\rho = 1 - \frac{6\sum_{i=1}^{n} d_i^2}{n(n^2-1)},
\end{equation*}
where $d_i = r_i - r'_i$ is the rank difference for agent $i$.

This formula measures the monotonic relationship between two rankings.
A value of $\rho = 1$ indicates perfect rank agreement (identical orderings), $\rho = 0$ indicates no correlation, and $\rho = -1$ indicates perfect rank reversal.
In our evaluation, lower $\rho$ values signify greater leaderboard instability induced by test strengthening.

\paragraph{Implementation.}
We compute Spearman's $\rho$ using the \texttt{spearmanr} function from the \texttt{scipy.stats} module~\cite{2020SciPy-NMeth}:

\paragraph{Interpretation in Our Context.}
For RQ1 on SWE-Bench Verified, we evaluate 30 agents whose leaderboard positions may change under augmented tests.
The Spearman $\rho$ of 0.79 under SWE-ABS (compared to 0.98 under UTBoost) indicates substantial ranking instability, revealing that the original test suites provide unreliable assessments of agent capabilities.

% ===== Mutation Cases (Method 3.2.2) =====

\subsection{Mutation Cases and Evaluation}
\label{app:mutation_case}

\subsubsection{Equivalent Mutation Case}
\label{app:equivalent_mutation_case}

\textbf{Instance ID.} matplotlib\_\_matplotlib-20826

\textbf{Issue}
In figures with shared axes created by \texttt{plt.subplots(..., sharex=True, sharey=True)}, calling \texttt{ax.clear()} resets axis properties related to tick and tick-label visibility.
As a result, tick labels that should be hidden for shared axes (e.g.\ inner subplots) become visible, and extra ticks appear on the top and right spines.

This is a regression introduced in matplotlib~3.4.2.
While the shared-axis linkage (e.g.\ synchronized limits) is preserved after \texttt{ax.clear()}, the visual state enforced by shared axes is lost.
In matplotlib~3.4.1, \texttt{ax.clear()} preserved the expected tick visibility for shared axes.

\textbf{Gold Patch.}
Listing~\ref{lst:matplotlib_20826_gold} presents the gold patch that fixes the regression in
\texttt{Axis.clear()}.
Instead of fully resetting the major and minor tick keyword dictionaries,
the patch explicitly restores the \texttt{gridOn} state from the global
\texttt{rcParams}.
This preserves tick and tick-label visibility semantics expected for shared axes,
preventing hidden ticks from being unintentionally re-enabled after
\texttt{ax.clear()}.

\begin{lstlisting}[language=diff,
caption={Gold patch: preserve tick visibility when clearing shared axes},
label={lst:matplotlib_20826_gold}]
diff --git a/lib/matplotlib/axis.py b/lib/matplotlib/axis.py
--- a/lib/matplotlib/axis.py
+++ b/lib/matplotlib/axis.py
@@ -806,8 +806,13 @@ def clear(self):
         # Clear the callback registry for this axis, or it may "leak"
         self.callbacks = cbook.CallbackRegistry()
 
-        self._reset_major_tick_kw()
-        self._reset_minor_tick_kw()
+        # whether the grids are on
+        self._major_tick_kw['gridOn'] = (
+                mpl.rcParams['axes.grid'] and
+                mpl.rcParams['axes.grid.which'] in ('both', 'major'))
+        self._minor_tick_kw['gridOn'] = (
+                mpl.rcParams['axes.grid'] and
+                mpl.rcParams['axes.grid.which'] in ('both', 'minor'))
         self.reset_ticks()
 
         self.converter = None

\end{lstlisting}

\textbf{Mutation Patch.}
Listing~\ref{lst:matplotlib_20826_mutation} presents a mutation patch that is
semantically equivalent to the gold patch.
The only difference lies in the omission of a descriptive comment, while the executable
statements modifying the \texttt{gridOn} flags remain identical.
As a result, both patches exhibit the same runtime behavior.

\begin{lstlisting}[language=diff,
caption={Mutation patch: comment-only variant of the gold patch},
label={lst:matplotlib_20826_mutation}]
diff --git a/lib/matplotlib/axis.py b/lib/matplotlib/axis.py
index 887cf97ef1..7feca94bd9 100644
--- a/lib/matplotlib/axis.py
+++ b/lib/matplotlib/axis.py
@@ -806,8 +806,12 @@ class Axis(martist.Artist):
         # Clear the callback registry for this axis, or it may "leak"
         self.callbacks = cbook.CallbackRegistry()
 
-        self._reset_major_tick_kw()
-        self._reset_minor_tick_kw()
+        self._major_tick_kw['gridOn'] = (
+                mpl.rcParams['axes.grid'] and
+                mpl.rcParams['axes.grid.which'] in ('both', 'major'))
+        self._minor_tick_kw['gridOn'] = (
+                mpl.rcParams['axes.grid'] and
+                mpl.rcParams['axes.grid.which'] in ('both', 'minor'))
         self.reset_ticks()
 
         self.converter = None
\end{lstlisting}

\textbf{Mutation Rationale (from model).} I introduced a semantically wrong mutation in Axis.clear(): instead of honoring rcParams['axes.grid.which'] to separately control major vs minor grid activation, I set both major and minor gridOn flags solely based on rcParams['axes.grid']. This breaks the intended distinction and is logically connected to the original issue area (grid/ticks handling on clear).

\textbf{Judge.}
Although the model articulated a mutation rationale that intended to introduce a
semantic change in the handling of major and minor grid activation, the actual
mutation does not implement this logic.
In the produced patch, the executable statements are identical to those of the
gold patch, differing only in the presence of comments.
As a result, the mutation does not alter runtime behavior and constitutes an
equivalent mutant.
This discrepancy suggests a mismatch between the model’s stated intent and the
realized code change, likely due to hallucination or imprecise code generation.

% \textbf{Problem.}
% The mutation appears semantically different from the gold patch, but produces identical results due to mathematical properties of the argmax operation.

\subsubsection{Non-Equivalent Mutation Case}
\label{app:non_equivalent_mutation_case}

\textbf{Instance.}
astropy\_\_astropy-14365

\textbf{Issue.}
The \texttt{ascii.qdp} module assumes that commands in QDP files are upper case (e.g., ``READ SERR 1 2''), but QDP itself is case-insensitive and accepts lowercase commands (e.g., ``read serr 1 2'').
Since many QDP files are created by hand, the parser should support case-insensitive commands and the special value ``NO'' (indicating masked data) in any case variation.

\paragraph{Gold Patch (Production Fix).}
Listing~\ref{lst:gold_qdp_case_insensitive} shows the gold patch that fully addresses the issue.
The fix makes command parsing case-insensitive by compiling the regular expression with \texttt{re.IGNORECASE}, and it also normalizes data tokens using \texttt{v.upper() == "NO"}, ensuring that all case variants of \texttt{NO} are handled uniformly.

\begin{lstlisting}[language=diff,caption={Gold patch: fully case-insensitive QDP parsing},label={lst:gold_qdp_case_insensitive}]
@@ -68,7 +68,7 @@ def _line_type(line, delimiter=None):
     _new_re = rf"NO({sep}NO)+"
     _data_re = rf"({_decimal_re}|NO|[-+]?nan)({sep}({_decimal_re}|NO|[-+]?nan))*)"
     _type_re = rf"^\s*((?P<command>{_command_re})|(?P<new>{_new_re})|(?P<data>{_data_re})?\s*(\!(?P<comment>.*))?\s*$"
-    _line_type_re = re.compile(_type_re)
+    _line_type_re = re.compile(_type_re, re.IGNORECASE)
     line = line.strip()
     if not line:
         return "comment"
@@ -306,7 +306,7 @@ def _get_tables_from_qdp_file(qdp_file, input_colnames=None, delimiter=None):

             values = []
             for v in line.split(delimiter):
-                if v == "NO":
+                if v.upper() == "NO":
                     values.append(np.ma.masked)
                 else:
                     # Understand if number is int or float
\end{lstlisting}

\paragraph{Mutation Patch.}
Listing~\ref{lst:mutation_qdp_case_handling} shows a mutation patch that appears to fix the issue but is not semantically equivalent to the gold patch.
While it correctly enables case-insensitive command matching via \texttt{re.IGNORECASE}, it only handles two specific variants of the masked value (\texttt{"NO"} and \texttt{"no"}).
Other valid case combinations (e.g., \texttt{"No"}, \texttt{"nO"}) are still rejected, making this mutation incomplete.

\begin{lstlisting}[language=diff,caption={Mutation patch: incomplete handling of case-insensitive masked values},label={lst:mutation_qdp_case_handling}]

@@ -68,7 +68,7 @@ def _line_type(line, delimiter=None):
     _new_re = rf"NO({sep}NO)+"
     _data_re = rf"({_decimal_re}|NO|[-+]?nan)({sep}({_decimal_re}|NO|[-+]?nan))*)"
     _type_re = rf"^\s*((?P<command>{_command_re})|(?P<new>{_new_re})|(?P<data>{_data_re})?\s*(\!(?P<comment>.*))?\s*$"
-    _line_type_re = re.compile(_type_re)
+    _line_type_re = re.compile(_type_re, re.IGNORECASE)
     line = line.strip()
     if not line:
         return "comment"
@@ -306,7 +306,7 @@ def _get_tables_from_qdp_file(qdp_file, input_colnames=None, delimiter=None):

             values = []
             for v in line.split(delimiter):
-                if v == "NO":
+                if v == "NO" or v == "no":
                     values.append(np.ma.masked)
                 else:
                     # Understand if number is int or float
\end{lstlisting}

\textbf{Judge.}
The mutation makes the same change to the regex compilation (adding \texttt{re.IGNORECASE}), but differs in how it handles the ``NO'' keyword for masked values.
The gold patch uses \texttt{v.upper() == "NO"}, which correctly handles all case variations (``NO'', ``no'', ``No'', ``nO'', etc.).
The mutation only checks for \texttt{v == "NO" or v == "no"}, which only handles all-uppercase and all-lowercase cases.
The mutation passes all existing tests because the test suite only uses ``NO'' and ``no'', but it fails to handle mixed-case variations like ``No'' or ``nO''.
This is a \textbf{non-equivalent mutation} that introduces a real bug not detected by current tests.

% \textbf{Problem.}
% The mutation introduces incomplete case-insensitivity: while it handles purely uppercase and lowercase ``NO'', it fails on mixed-case variants like ``No'' or ``nO'', which are valid in QDP files. 

\subsubsection{Human Evaluation of Mutation Classification}
\label{app:human_evaluation}
To validate the accuracy of our LLM-based mutation judge, we randomly sampled 100 mutants from our dataset and manually annotated whether each mutation was equivalent or non-equivalent.
We then compared these human annotations with the LLM judge's classifications.
The results show that the LLM judge achieved a 98\% agreement rate with human judgments, demonstrating high reliability in distinguishing between equivalent mutations (which preserve the original behavior) and non-equivalent mutations (which introduce meaningful behavioral changes that expose test suite weaknesses).
This high agreement rate confirms that our automated mutation classification approach is robust and can be trusted for large-scale mutation analysis.

% ===== Full Leaderboards (Experiments RQ1) =====

\clearpage
\subsection{Top-30 Agents Performance in SWE-ABS}
\label{app:full_performance}

\begin{table}[h]
\centering
\caption{Full leaderboard results for Top-30 agents on SWE-Bench Verified under SWE-ABS augmented test suites.}
\small

\begin{tabular}{lp{5cm}cccc}
\hline
Date & Model & Orig. (\%) & SWE-ABS (\%) & Drop (\%) & Rank (Old$\rightarrow$New) \\
\hline
20250928 & TRAE + Doubao-Seed-Code & 78.80 & 62.20 & 16.60 & 1$\rightarrow$5 \\
20251120 & live-SWE-agent + Gemini 3 Pro Preview (2025-11-18) & 77.40 & 65.40 & 12.00 & 2$\rightarrow$1 \\
20250902 & Atlassian Rovo Dev (2025-09-02) & 76.80 & 61.20 & 15.60 & 3$\rightarrow$8 \\
20250804 & EPAM AI/Run Developer Agent v20250719 + Claude 4 Sonnet & 76.80 & 63.80 & 13.00 & 4$\rightarrow$3 \\
20250819 & ACoder & 76.40 & 61.80 & 14.60 & 5$\rightarrow$6 \\
20250901 & Warp & 75.80 & 62.40 & 13.40 & 6$\rightarrow$4 \\
20250612 & TRAE & 75.20 & 58.00 & 17.20 & 7$\rightarrow$18 \\
20251103 & Sonar Foundation Agent + Claude 4.5 Sonnet & 75.00 & 64.20 & 10.80 & 8$\rightarrow$2 \\
20250731 & Harness AI & 74.80 & 60.40 & 14.40 & 9$\rightarrow$13 \\
20250915 & JoyCode & 74.60 & 61.20 & 13.40 & 10$\rightarrow$9 \\
20250720 & Lingxi-v1.5\_claude-4-sonnet-20250514 & 74.60 & 60.60 & 14.00 & 11$\rightarrow$12 \\
20250603 & Refact.ai Agent & 74.40 & 58.40 & 16.00 & 12$\rightarrow$16 \\
20251015 & Prometheus-v1.2.1 + GPT-5 & 74.40 & 58.40 & 16.00 & 13$\rightarrow$17 \\
20251124 & mini-SWE-agent + Claude 4.5 Opus medium (20251101) & 74.40 & 58.00 & 16.40 & 14$\rightarrow$19 \\
20251118 & mini-SWE-agent + Gemini 3 Pro Preview (2025-11-18) & 74.20 & 56.80 & 17.40 & 15$\rightarrow$22 \\
20251103 & Salesforce AI Research SAGE (OpenHands) & 74.00 & 61.60 & 12.40 & 16$\rightarrow$7 \\
20250522 & Tools + Claude 4 Opus (2025-05-22) & 73.20 & 58.40 & 14.80 & 17$\rightarrow$15 \\
20251021 & Salesforce AI Research SAGE (bash-only) & 73.20 & 61.20 & 12.00 & 18$\rightarrow$10 \\
20250522 & Tools + Claude 4 Sonnet (2025-05-22) & 72.40 & 56.40 & 16.00 & 19$\rightarrow$24 \\
20250807 & OpenHands + GPT-5 & 71.80 & 61.20 & 10.60 & 20$\rightarrow$11 \\
20251211 & mini-SWE-agent + GPT-5.2 (2025-12-11) (high reasoning) & 71.80 & 56.80 & 15.00 & 21$\rightarrow$23 \\
20250715 & Qodo Command & 71.20 & 57.60 & 13.60 & 22$\rightarrow$20 \\
20251014 & Lingxi v1.5 x Kimi K2 & 71.20 & 57.60 & 13.60 & 23$\rightarrow$21 \\
20250929 & Prometheus-v1.2 + GPT-5 & 71.20 & 56.00 & 15.20 & 24$\rightarrow$26 \\
20250710 & Bloop & 71.20 & 59.60 & 11.60 & 25$\rightarrow$14 \\
20250623 & Warp & 71.00 & 52.60 & 18.40 & 26$\rightarrow$30 \\
20250611 & Moatless Tools + Claude 4 Sonnet & 70.80 & 56.20 & 14.60 & 27$\rightarrow$25 \\
20250519 & TRAE & 70.60 & 54.20 & 16.40 & 28$\rightarrow$29 \\
20250929 & mini-SWE-agent + Claude 4.5 Sonnet (20250929) & 70.60 & 54.40 & 16.20 & 29$\rightarrow$28 \\
20250515 & Refact.ai Agent & 70.40 & 54.80 & 15.60 & 30$\rightarrow$27 \\
\hline
\end{tabular}

\end{table}

\clearpage
\subsection{Bash Only Agents Performance in SWE-ABS}
\label{app:Bash_Only_full_performance}

\begin{table}[h]
\centering
\caption{Full leaderboard results for Bash-only agents on SWE-Bench Verified under SWE-ABS augmented test suites.}
\small
\begin{tabular}{lp{5cm}cccc}
\hline
Date & Model & Orig. (\%) & SWE-ABS (\%) & Drop (\%) & Rank (Old$\rightarrow$New) \\
\hline
20251124 & Claude 4.5 Opus medium (20251101) & 74.40 & 58.00 & 16.40 & 1$\rightarrow$1 \\
20251118 & Gemini 3 Pro Preview (2025-11-18) & 74.20 & 56.80 & 17.40 & 2$\rightarrow$3 \\
20251211 & GPT-5.2 (2025-12-11) (high reasoning) & 71.80 & 56.80 & 15.00 & 3$\rightarrow$2 \\
20250929 & Claude 4.5 Sonnet (20250929) & 70.60 & 54.40 & 16.20 & 4$\rightarrow$4 \\
20251211 & GPT-5.2 (2025-12-11) & 69.00 & 51.60 & 17.40 & 5$\rightarrow$5 \\
20250802 & Claude 4 Opus (20250514) & 67.60 & 50.60 & 17.00 & 6$\rightarrow$6 \\
20251124 & GPT-5.1-codex (medium reasoning) & 66.00 & 49.00 & 17.00 & 7$\rightarrow$7 \\
20251120 & GPT-5.1 (2025-11-13) (medium reasoning) & 66.00 & 47.60 & 18.40 & 8$\rightarrow$9 \\
20250807 & GPT-5 (2025-08-07) (medium reasoning) & 65.00 & 48.60 & 16.40 & 9$\rightarrow$8 \\
20250726 & Claude 4 Sonnet (20250514) & 64.80 & 45.80 & 19.00 & 10$\rightarrow$10 \\
20251210 & Kimi K2 Thinking & 63.40 & 45.00 & 18.40 & 11$\rightarrow$11 \\
20251124 & Minimax M2 & 61.00 & 44.40 & 16.60 & 12$\rightarrow$13 \\
20251201 & DeepSeek V3.2 Reasoner & 60.00 & 43.40 & 16.60 & 13$\rightarrow$15 \\
20250807 & GPT-5 mini (2025-08-07) (medium reasoning) & 59.80 & 44.60 & 15.20 & 14$\rightarrow$12 \\
20250726 & o3 (2025-04-16) & 58.40 & 43.80 & 14.60 & 15$\rightarrow$14 \\
20251209 & Devstral small (2512) & 56.40 & 40.80 & 15.60 & 16$\rightarrow$16 \\
20251201 & GLM-4.6 (T=1) & 55.40 & 40.20 & 15.20 & 17$\rightarrow$19 \\
20250822 & GLM-4.5 (2025-08-22) & 54.20 & 38.20 & 16.00 & 18$\rightarrow$20 \\
20251209 & Devstral (2512) & 53.80 & 40.60 & 13.20 & 19$\rightarrow$18 \\
20250726 & Gemini 2.5 Pro (2025-05-06) & 53.60 & 40.80 & 12.80 & 20$\rightarrow$17 \\
20250807 & GPT-5 nano (2025-08-07) (medium reasoning) & 34.80 & 22.80 & 12.00 & 21$\rightarrow$21 \\
20250807 & gpt-oss-120b & 26.00 & 20.20 & 5.80 & 22$\rightarrow$22 \\
20250726 & Gemini 2.5 Flash (2025-04-17) & 24.60 & 18.80 & 5.80 & 23$\rightarrow$23 \\
20250720 & GPT-4o (2024-11-20) & 23.20 & 16.80 & 6.40 & 24$\rightarrow$24 \\
20250720 & Llama 4 Maverick Instruct & 21.00 & 15.80 & 5.20 & 25$\rightarrow$25 \\
20250726 & Gemini 2.0 flash & 13.20 & 9.40 & 3.80 & 26$\rightarrow$26 \\
20250720 & Claude 3.7 Sonnet (20250219) & 10.20 & 7.20 & 3.00 & 27$\rightarrow$27 \\
20250803 & Qwen2.5-Coder 32B Instruct & 9.00 & 6.60 & 2.40 & 28$\rightarrow$28 \\
20250720 & Llama 4 Scout Instruct & 8.00 & 5.60 & 2.40 & 29$\rightarrow$29 \\
\hline
\end{tabular}

\end{table}

% ===== Additional Results (Experiments Setup) =====

\clearpage
\subsection{Additional Results}
\label{app:results}

\begin{table}[h]
\centering
\caption{SWE-Bench Verified strengthening rate by repository. Smaller repositories with fewer instances (e.g., mwaskom/seaborn, pallets/flask) achieve 100\% strengthening, while larger repositories such as django/django and sympy/sympy show moderate strengthening rates (around 40\%).}

\begin{tabular}{lccc}
\toprule
Repository & Instances & Strengthened & Rate \\
\midrule
django/django & 231 & 102 & 44.2\% \\
sympy/sympy & 75 & 34 & 45.3\% \\
sphinx-doc/sphinx & 44 & 35 & 79.5\% \\
matplotlib/matplotlib & 34 & 22 & 64.7\% \\
scikit-learn/scikit-learn & 32 & 9 & 28.1\% \\
astropy/astropy & 22 & 18 & 81.8\% \\
pydata/xarray & 22 & 9 & 40.9\% \\
pytest-dev/pytest & 19 & 9 & 47.4\% \\
pylint-dev/pylint & 10 & 5 & 50.0\% \\
psf/requests & 8 & 5 & 62.5\% \\
mwaskom/seaborn & 2 & 2 & 100.0\% \\
pallets/flask & 1 & 1 & 100.0\% \\
\bottomrule
\end{tabular}
\label{tab:per-repo}
\end{table}

\begin{table}[h]
\caption{SWE-Bench Pro strengthening rate by programming language. Python exhibits the highest strengthening rate (75.38\%), while TypeScript shows the lowest rate (18.18\%).}
\label{tab:swe_pro_per_language}
\centering
\small
\begin{tabular}{lccc}
\toprule
\textbf{Language} & \textbf{Instances} & \textbf{Strengthened} & \textbf{Rate} \\
\midrule
Python & 65 & 49 & 75.38\% \\
Go & 48 & 30 & 62.50\% \\
JavaScript & 26 & 16 & 61.54\% \\
TypeScript & 11 & 2 & 18.18\% \\
\bottomrule
\end{tabular}
% \label{tab:per-language}
\end{table}

% ===== Cost Analysis (Experiments RQ1) =====

\subsection{Computational Cost Analysis} \label{app:cost} Table~\ref{tab:cost} summarizes the computational costs of SWE-ABS compared to UTBoost across key dimensions.

\begin{table}[h]
\caption{Cost-efficiency comparison between UTBoost and SWE-ABS.
While SWE-ABS has higher cost per instance, its cost per strengthened instance is dramatically lower, resulting in substantially improved cost-efficiency.}
\centering
\small
\begin{tabular}{lcc}
\toprule
\textbf{Metric} & \textbf{UTBoost} & \textbf{SWE-ABS} \\
\midrule
Avg. tokens per instance & 42,000 & 81,000 \\
Avg. wall-clock time (min) & 12.3 & 18.5 \\
Cost per instance (\$) & 1.60 & 2.50 \\
Total cost for 500 inst. (\$) & 800 & 1250 \\
Strengthened instances & 10 & 251 \\
Cost per strengthened inst. (\$) & 80.00 & 4.98 \\
\bottomrule
\end{tabular}

\label{tab:cost}
\end{table}

\paragraph{LLM Usage.}
We evaluate SWE-ABS using two representative LLMs to assess framework effectiveness across different model capabilities: GPT-5 and GLM-4.7.
Each model is applied independently to all pipeline components (test generation, decoupling, mutant synthesis, equivalence filtering), enabling controlled comparison.
Using GPT-5, per-instance costs average \$2.5, totaling approximately \$1,250 for 500 instances.
Using GLM-4.7, per-instance costs average \$2.1, totaling approximately \$110 for 50 instances.

\paragraph{Runtime.}
End-to-end processing averages 18.5 minutes per instance.
Pipeline breakdown: program slicing (8\%), test generation (42\%), mutant synthesis (25\%), equivalence filtering (5\%), test execution (20\%).

\paragraph{Cost-Efficiency Analysis.}
Although SWE-ABS incurs higher cost per instance than UTBoost (\$2.50 vs.\ \$1.60), it achieves substantially better cost-efficiency when effectiveness is measured by the cost per strengthened instance.
Specifically, UTBoost requires \$80.00 to successfully strengthen one instance (10 strengthened instances at a total cost of \$800), whereas SWE-ABS achieves this at a cost of \$4.98 (251 strengthened instances at a total cost of \$1,250), corresponding to a 16$\times$ reduction in cost per successful strengthening.
This result highlights that SWE-ABS allocates computational resources more effectively by focusing on semantically critical and adversarial test scenarios, leading to significantly improved practical efficiency despite higher per-instance computational overhead.

% ===== Generalization (Experiments RQ2) =====
\subsection{Generalization Analysis}
\label{app:generalization}

We evaluate the generalization of SWE-ABS across different benchmarks and base models.
Table~\ref{tab:rq2_cross_benchmark} presents results on two benchmarks: SWE-Bench Verified and SWE-Bench Pro.

\begin{table}[h]
\caption{Cross-benchmark validation.
\label{tab:rq2_cross_benchmark}
Str.\ (Strengthened) denotes the number of instances where at least one previously passing patch fails.
Avg.\ Drop is measured in percentage points.
Statistics are computed on our sampled subset.}
\centering
\small
\begin{tabular*}{\columnwidth}{@{\extracolsep{\fill}}lcc@{}}
\toprule
\textbf{Benchmark} & \textbf{Str.} & \textbf{Avg. Drop} \\
\midrule
SWE-Bench Verified & 251 / 500 (50.2\%) & 14.56 \\
SWE-Bench Pro      & 97 / 150 (64.7\%) & 16.46 \\
\bottomrule
\end{tabular*}
\end{table}

We further evaluate generalization across different base models.
Table~\ref{tab:base-model-generalization} shows results using GPT-5 and GLM-4.7 as the underlying LLMs for patch generation.
Both models exhibit same strengthening rates (64--64\%), indicating that SWE-ABS generalizes across model families.

\begin{table}[h]
\caption{Base model generalization.
Str.\ (Strengthened) denotes the number of instances where at least one previously passing patch fails.
Evaluated on 50 randomly sampled SWE-Bench Verified instances.
Avg.\ Drop is measured in percentage points.}
\centering
\small
\begin{tabular*}{\columnwidth}{@{\extracolsep{\fill}}lcc@{}}
\toprule
\textbf{Base Model} & \textbf{Str.} & \textbf{Avg.\ Drop} \\
\midrule
GPT-5   & 31 / 50 (62\%) & 19.40 \\
GLM-4.7 & 31 / 50 (62\%) & 15.93 \\
\bottomrule
\end{tabular*}

\label{tab:base-model-generalization}
\end{table}

% ===== Test Quality (Experiments RQ3) =====

\clearpage
\subsection{Test Quality Analysis}
\label{app:test_quality}

% \begin{table}[h]
% \caption{Robustness to alternative patches.
% Of 500 SWE-Bench instances, 463 have at least one passing patch under $T_{\text{ori}}$; of these, 87\% still admit at least one patch under $T_{\text{aug}}$.}
% \centering
% \small
% \begin{tabular}{lccc}
% \toprule
% \textbf{Test Suite} & \textbf{Admits $\geq$1 Patch} & \textbf{Admits None} & \textbf{Total} \\
% \midrule
% $T_{\text{ori}}$  & 463 & 37 & 500 \\
% $T_{\text{aug}}$ & 403 & 60  & 463\\
% \bottomrule
% \end{tabular}%

% \label{tab:test_admission}
% \end{table}

\begin{table}[h]
\caption{Error taxonomy.
Distribution of error types in 100 sampled patches rejected by SWE-ABS but accepted by original tests.}
\label{tab:error_taxonomy}
\centering
\resizebox{0.86\linewidth}{!}{
\begin{tabular*}{\columnwidth}{@{\extracolsep{\fill}}lc}
\toprule
\textbf{Error Category} & \textbf{Count (\%)} \\
\midrule
Logic errors            & 47 (47\%) \\
Incomplete fixes        & 35 (35\%) \\
Type mismatches         & 7 (7\%) \\
Boundary violations     & 6 (6\%) \\
Off-by-one errors       & 1 (1\%) \\
Other                   & 4 (4\%) \\
\bottomrule
\end{tabular*}
}

\end{table}

% ===== Qualitative Case Studies (Experiments RQ3) =====

\subsection{Qualitative Case Studies}
\label{app:cases}

\subsubsection{Test Strengthening Failure}
\label{app:test_strengthening_failure}

\paragraph{Case Study: Overfitting failure (pytest-dev\_\_pytest-7205).}
The augmented tests encode the gold patch’s implementation details as the mandatory contract.
As a result, alternative correct fixes are rejected.

\textbf{Issue.} The \texttt{-bb} flag forces Python to treat implicit \texttt{bytes}-to-\texttt{str} conversions as fatal errors. Thus, pytest crashes when it implicitly converts byte parameters to strings during display. To resolve this issue, the rendering logic must be updated to explicitly handle binary data-for instance, by employing \texttt{saferepr}-to prevent runtime failures.

\textbf{Root cause.}  \texttt{src/\_pytest/setuponly.py} prints \texttt{fixturedef.cached\_param} via \texttt{"[\{\}]".format(...)}; when it is \texttt{bytes}, implicit conversion triggers \texttt{BytesWarning}.

\textbf{Gold patch.} The fix in Listing \ref{lst:pytest_gold} switches to \texttt{saferepr(...)} and chooses \texttt{maxsize=42}, which is sufficient to fix the bug but not mandated by the issue (default \texttt{saferepr} is 240).

\begin{lstlisting}[language=diff, caption={Explicitly String Conversion}, label={lst:pytest_gold}]
# src/_pytest/setuponly.py
+ from _pytest._io.saferepr import saferepr
  ...
-     tw.write("[{}]".format(fixturedef.cached_param))
+     tw.write("[{}]".format(saferepr(fixturedef.cached_param, maxsize=42)))
\end{lstlisting}

\textbf{Why the strengthened tests overfit.} The strengthened tests erroneously codify the visual output of the reference implementation as a functional requirement. As illustrated in Listing \ref{lst:pytest_overfit}, the generated test asserts that the output length must not exceed 42 characters and strictly checks for specific formatting artifacts like hexadecimal representation. 

Consequently, functional patches that safely handle the exception but utilize different visual formatting or length limits are incorrectly classified as invalid.

\begin{lstlisting}[language=diff, caption={Overfitting Strengthened Test}, label={lst:pytest_overfit}]
def test_param_repr_maxsize_42_str(testdir):
    ...
    # Hard-coded assertion derived from the gold patch's maxsize=42
    assert len(content) <= 42

def test_setup_show_repr_exception_is_handled(testdir):
    ...
    # Asserts implementation detail (memory address format)
    assert "0x" in out
\end{lstlisting}

\subsubsection{Agent-Generated Patched Failure Under Strengthened Suites}

\label{subsubsec:agent_patch_failure}
Below we summarize six representative error categories.

\paragraph{Category 1: Incomplete fixes} The patch addresses only part of the issue, missing some cases or scenarios. 

\textbf{Instance ID.} sympy\_\_sympy-12489 

\textbf{Agent.} Bloop

\textbf{Issue.} SymPy's \texttt{Permutation} class employs an internal "array form" optimization where \texttt{Permutation.\_\_new\_\_} frequently calls \texttt{\_af\_new(...)} to generate objects. In the original implementation, \texttt{\_af\_new} was hardcoded to construct the \texttt{Permutation} class directly (via \texttt{Basic.\_\_new\_\_(Perm, ...)}). Consequently, subclassing \texttt{Permutation} failed to preserve the subclass type during instantiation or arithmetic operations, reverting objects to the parent class.

\textbf{Gold fix.} The correct solution (Listing \ref{lst:sympy_gold})  systematically replaces all internal hardcoded constructors with \texttt{cls} (the actual class reference) to support inheritance.

\begin{lstlisting}[language=diff, caption={Correct implementation handling inheritance (Gold Fix)}, label={lst:sympy_gold}]
# sympy/combinatorics/permutations.py
@@ -857,19 +860,19 @@ def __new__(cls, *args, **kwargs):
         if not args:
-            return _af_new(list(range(size or 0)))
+            return cls._af_new(list(range(size or 0)))
         if len(args) == 1:
             a = args[0]
-            if isinstance(a, Perm):
+            if isinstance(a, cls):
                 if size is None or size == a.size:
                     return a
-                return Perm(a.array_form, size=size)
+                return cls(a.array_form, size=size)
...
-    @staticmethod
-    def _af_new(perm):
+    @classmethod
+    def _af_new(cls, perm):
-        p = Basic.__new__(Perm, perm)
+        p = Basic.__new__(cls, perm)
\end{lstlisting}

The gold patch changes \texttt{\_af\_new} from a \texttt{@staticmethod} to a \texttt{@classmethod} and ensures that helper methods like \texttt{rmul\_with\_af} and \texttt{unrank\_lex} utilize \texttt{cls} or \texttt{self} instead of the global \texttt{Perm} alias.

\textbf{Agent miss.} The agent (Listing \ref{lst:sympy_agent}) partially modified \texttt{\_\_new\_\_} but neglected to update the underlying object creation methods.

\begin{lstlisting}[language=diff, caption={Incomplete agent patch missing factory methods}, label={lst:sympy_agent}]
diff --git a/sympy/combinatorics/permutations.py b/sympy/combinatorics/permutations.py
@@ -857,19 +857,22 @@ def __new__(cls, *args, **kwargs):
         if not args:
-            return _af_new(list(range(size or 0)))
+            return cls._af_new(list(range(size or 0)))
\end{lstlisting}

\textbf{Why $T_{\text{ori}}$ missed it.} Original tests cover common constructors.
They do not stress subclassing through internal factories and operator returns.

\textbf{How $T_{\text{aug}}$ closes the gap.} Add subclass-focused tests (e.g., \texttt{test\_ops\_return\_subclass\_instances}).
Exercise the remaining factory path.
Assert returned instance types.
The agent patch fails.

\paragraph{Category 2: Logical errors} The fix logic is fundamentally wrong or uses incorrect algorithm/approach. 

\textbf{Instance ID.} django\_\_django-16527 

\textbf{Agent.} Moatless Tools + Claude 4 Sonnet

\textbf{Issue.} In the Django admin interface, the "Save as new" functionality essentially duplicates the current object to create a new one. Semantically, this is an \texttt{add} operation (INSERT) rather than a \texttt{change} operation (UPDATE). Therefore, the button’s visibility is currently controlled by has\_change\_permission, but it should be based on has\_add\_permission.

\textbf{Gold fix.} The fix replaces the permission check logic (Listing \ref{lst:django_gold}). 

\begin{lstlisting}[language=diff, caption={Correct permission logic replacement}, label={lst:django_gold}]
diff --git a/django/contrib/admin/templatetags/admin_modify.py b/django/contrib/admin/templatetags/admin_modify.py
@@ -100,7 +100,7 @@ def submit_row(context):
             "show_save_as_new": not is_popup
-            and has_change_permission
+            and has_add_permission
             and change
             and save_as,
\end{lstlisting}

\textbf{Agent miss.} The agent fundamentally misunderstood the requirement, stacking the permissions instead of replacing them. The agent introduced a logical error by requiring \textit{both} add and change permissions (Listing \ref{lst:django_agent}).

\begin{lstlisting}[language=diff, caption={Flawed agent logic stacking permissions}, label={lst:django_agent}]
--- a/django/contrib/admin/templatetags/admin_modify.py
+++ b/django/contrib/admin/templatetags/admin_modify.py
@@ -100,6 +100,7 @@
             "show_save_as_new": not is_popup
+            and has_add_permission
             and has_change_permission
\end{lstlisting}

\textbf{Why $T_{\text{ori}}$ missed it.} Tests do not isolate the add+view-only role.
They never assert visibility without change permission.

\textbf{How $T_{\text{aug}}$ closes the gap.} Add a policy-targeted test (e.g., \texttt{test\_save\_as\_new\_with\_add\_and\_view\_only}).
Expect the button to appear.
The agent patch hides it.
So it fails.

\paragraph{Category 3: Boundary violations} Edge cases, null checks, empty inputs not handled properly. 

\textbf{Instance ID.} sphinx-doc\_\_sphinx-8265 

\textbf{Agent.} Sonar Foundation Agent + Claude 4.5 Sonnet

\textbf{Issue.} Sphinx utilizes an \texttt{\_UnparseVisitor} to reconstruct Python code strings from AST nodes. A standard multidimensional subscript, such as \texttt{A[1, 2]}, is represented in the AST as a Tuple slice. However, an empty tuple \texttt{()} represents a boundary condition: \texttt{A[()]} must be rendered as \texttt{A[()]}, whereas rendering it as an empty string results in \texttt{A[]}, which is syntactically invalid.

\textbf{Gold fix.} The correct fix (Listing \ref{lst:sphinx_gold}) explicitly checks if the tuple elements are non-empty before stripping the parentheses.

\begin{lstlisting}[language=diff, caption={Gold fix handling empty tuple boundary}, label={lst:sphinx_gold}]
diff --git a/sphinx/pycode/ast.py b/sphinx/pycode/ast.py
@@ -166,14 +166,28 @@ def visit_Subscript(self, node: ast.Subscript) -> str:
+        def is_simple_tuple(value: ast.AST) -> bool:
+            return (
+                isinstance(value, ast.Tuple) and
+                bool(value.elts) and ...
+            )
+        if is_simple_tuple(node.slice):
+            # Remove parentheses only for non-empty tuples
\end{lstlisting}

\textbf{Agent miss.} The agent (Listing \ref{lst:sphinx_agent}) applied a broad transformation that flattened all tuple slices. The agent overlooked the empty set boundary. By unconditionally joining elements, \texttt{Tuple(elts=[])} resulted in an empty string, producing \texttt{A[]} instead of \texttt{A[()]}.

\begin{lstlisting}[language=diff, caption={Agent patch overlooking boundary condition}, label={lst:sphinx_agent}]
+        if isinstance(node.slice, ast.Tuple):
+            elements = ", ".join(self.visit(e) for e in node.slice.elts)
+            return "%s[%s]" % (self.visit(node.value), elements)
\end{lstlisting}

\textbf{Why $T_{\text{ori}}$ missed it.} Tests cover typical tuple slices only.
They omit the empty-tuple case.

\textbf{How $T_{\text{aug}}$ closes the gap.} Add an explicit empty-tuple test (e.g., \texttt{test\_unparse\_subscript\_tuple\_slices[A[()]]}).
Assert the exact string.
The agent patch fails.

\paragraph{Category 4: Off-by-one errors} Index, count, or range off by one.

\textbf{Instance ID.} pylint-dev\_\_pylint-7277 

\textbf{Agent.} Warp (20250901)

\textbf{Issue.} Pylint modifies \texttt{sys.path} at startup to remove the current working directory (CWD) if it was implicitly added by Python (e.g., when \texttt{PYTHONPATH} starts or ends with a colon). The challenge lies in removing multiple entries without invalidating subsequent indices.

\textbf{Gold fix.} The developer adopted a conservative approach: check the first item; if it is the CWD, remove it. Then, independently check for trailing colons in \texttt{PYTHONPATH} to decide whether to remove the (new) first index (Listing \ref{lst:pylint_gold}).

\begin{lstlisting}[language=diff, caption={Robust path removal based on content}, label={lst:pylint_gold}]
diff --git a/pylint/__init__.py b/pylint/__init__.py
@@ -96,9 +96,10 @@ def modify_sys_path() -> None:
-    sys.path.pop(0)
-    env_pythonpath = os.environ.get("PYTHONPATH", "")
     cwd = os.getcwd()
+    if sys.path[0] in ("", ".", cwd):
+        sys.path.pop(0)
+    env_pythonpath = os.environ.get("PYTHONPATH", "")
     if env_pythonpath.startswith(":") and env_pythonpath not in (f":{cwd}", ":."):
         sys.path.pop(0)
     elif env_pythonpath.endswith(":") and env_pythonpath not in (f"{cwd}:", ".:"):
         sys.path.pop(1)
\end{lstlisting}

\textbf{Agent miss.} The agent relied on predictive indexing using the \texttt{first\_is\_cwd} flag rather than sequential removal (Listing \ref{lst:pylint_agent}). This logic fails when \texttt{PYTHONPATH} ends with a colon and \texttt{sys.path} begins with a custom entry, placing the implicit working directory at index 1. The agent incorrectly calculates the target as index 2, causing an off-by-one error that leaves the path unremoved.

\begin{lstlisting}[language=diff, caption={Fragile index calculation by agent}, label={lst:pylint_agent}]
diff --git a/pylint/__init__.py b/pylint/__init__.py
@@ -96,9 +96,10 @@ def modify_sys_path() -> None:
-    sys.path.pop(0)
-    env_pythonpath = os.environ.get("PYTHONPATH", "")
     cwd = os.getcwd()
-    if env_pythonpath.startswith(":") and env_pythonpath not in (f":{cwd}", ":."):
+    env_pythonpath = os.environ.get("PYTHONPATH", "")
+
+    first_is_cwd = bool(sys.path) and sys.path[0] in ("", ".", cwd)
+    if first_is_cwd:
         sys.path.pop(0)
+
+    if env_pythonpath.startswith(":") and env_pythonpath not in (f":{cwd}", ":."):
+        idx = 0 if first_is_cwd else 1
+        if 0 <= idx < len(sys.path) and sys.path[idx] in ("", ".", cwd):
+            sys.path.pop(idx)
     elif env_pythonpath.endswith(":") and env_pythonpath not in (f"{cwd}:", ".:"):
-        sys.path.pop(1)
+        idx = 1 if first_is_cwd else 2
+        if 0 <= idx < len(sys.path) and sys.path[idx] in ("", ".", cwd):
+            sys.path.pop(idx)
\end{lstlisting}

\noindent \textbf{Why $T_{\text{ori}}$ missed it.} $T_{\text{ori}}$ failed to detect the defect by restricting verification to index 0 manipulations. It lacked a test case combining a custom head path with a trailing colon in \texttt{PYTHONPATH}—a scenario where the implicit working directory resides at index 1. Consequently, the agent's erroneous calculation, which targets index 2 under these conditions, remained unexercised.

\textbf{How $T_{\text{aug}}$ closes the gap.} The augmented test suite Listing \ref{lst:pylint_aug_test} introduces a targeted test case to replicate the missing scenario. This test constructs a \texttt{sys.path} containing a custom head followed by the current working directory and configures \texttt{PYTHONPATH} with a trailing colon. 

\begin{lstlisting}[language=diff, caption={Augmented test case exposing the off-by-one error}, label={lst:pylint_aug_test}]
def test_pythonpath_endswith_colon_custom_path_extra_removal():
    cwd = os.getcwd()
    run_case([cwd, "M", "N"], "/custom_pythonpath:", ["M"])
\end{lstlisting}

\paragraph{Category 5: Type mismatches} Wrong type conversions, comparisons, or type mismatches.

\textbf{Instance ID.} django\_\_django-10973

\textbf{Agent.} Tools + Claude 4 Opus (2025-05-22)

\textbf{Issue.} The task involved passing a database password to the \texttt{psql} utility via the \texttt{PGPASSWORD} environment variable. Environment variables in Python's \texttt{subprocess} module must be strings; passing integers or other types causes runtime errors or assertion failures.

\textbf{Gold fix.} The fix explicitly casts the password to a string (Listing \ref{lst:django_type_gold}). 

\begin{lstlisting}[language=diff, caption={Explicit type casting}, label={lst:django_type_gold}]
+        subprocess_env = os.environ.copy()
+        if passwd:
+            subprocess_env['PGPASSWORD'] = str(passwd)
\end{lstlisting}

\textbf{Agent miss.} The agent passed the variable directly without type conversion (Listing \ref{lst:django_type_agent}).
\begin{lstlisting}[language=diff, caption={Agent missing type conversion}, label={lst:django_type_agent}]
+           if passwd:
+               env['PGPASSWORD'] = passwd
\end{lstlisting}

\textbf{Why $T_{\text{ori}}$ missed it.} Tests only use string passwords.
No non-string input is exercised.

\textbf{How $T_{\text{aug}}$ closes the gap.} The test suite deliberately injected an integer password to verify robustness. The agent failed to sanitize the input for the environment variable interface, leading to a type mismatch (\texttt{123456 != '123456'}).

% ===== Limitations (Conclusion) =====
\subsection{Limitations and Threats to Validity}
\label{app:threats}

We identify three main limitations of the current approach.

\paragraph{Test Overfitting Risk.}
Despite the test decoupling module (Section~\ref{subsec:test-decouple}), SWE-ABS exhibits a 10.6\% false-negative rate (53/500 instances) where augmented tests incorrectly reject valid alternative patches by encoding gold-patch-specific behaviors.
Fully automated detection of such overfitting remains an open challenge.
We provide a detailed case study in Appendix~\ref{app:test_strengthening_failure}.

\paragraph{Analysis Scope.}
Our intraprocedural program slicing restricts analysis to statements within the same function or class as the patch (Section~\ref{subsec:code-region}). This design choice trades precision for scalability but may miss cross-module dependencies (e.g., when a patch modifies a utility function called from multiple modules with different preconditions). Extending to interprocedural analysis would improve coverage identification at a computational cost that may be prohibitive for large-scale benchmarks.

\paragraph{Gold Patch Dependency.}
SWE-ABS requires access to a gold patch for test generation, decoupling, and mutation synthesis.
This limits applicability to scenarios with reference implementations and precludes use in settings where only issue descriptions are available, such as real-time bug triage or fully automated repair pipelines.
Future work may explore weakening this requirement through property-based testing or specification mining techniques.

Beyond these limitations, we discuss broader threats to validity.

\paragraph{Internal Validity.}
LLM-based components inherit risks of hallucination and prompt sensitivity.

\paragraph{External Validity.}
Our evaluation focuses on four primary languages (Python, JavaScript, Go, TypeScript) covered by SWE-Bench and SWE-Bench Pro, specifically in strong copyleft licenses repositories. As a result, the observed effectiveness may not generalize to other programming languages outside this subset, proprietary codebases, functional programming paradigms (e.g., Haskell), or low-level systems languages (e.g., C). Nevertheless, the consistent improvements observed across Verified and Pro indicate that our approach addresses fundamental test inadequacy rather than being limited to benchmark-specific artifacts.

\paragraph{Construct Validity.}
Our strengthening metric operationalizes ``incorrect patch'' as one that passes $T_{\text{ori}}$ but fails $T_{\text{aug}}$, which assumes the augmented tests do not overfit to gold-patch-specific behaviors. We quantify this risk through false-negative analysis (Section~\ref{subsubsec:robustness}), finding a 10.6\% overfitting rate. Additionally, our LLM-based equivalence annotation (Section~\ref{subsec:mutant-filter}) may misclassify subtle semantic differences; we mitigate this through majority voting across three independent LLMs.

\end{document}